\documentclass[singlecolumn,superscriptaddress,showkeys,showpacs,longbibliography,groupedaddress]{revtex4-1}
\usepackage{float}

\usepackage[utf8]{inputenc}
\usepackage{cmap} 
\usepackage[T1]{fontenc}
\usepackage{microtype}
\usepackage{amsmath,amssymb,mathtools}
\usepackage{txfonts}

\usepackage{appendix}
\usepackage{mathrsfs}
\usepackage{graphicx,grffile,textcomp}
\graphicspath{.}
\usepackage{textcomp}
\usepackage{booktabs,tabularx,dcolumn}
\usepackage{array}
\newcolumntype{P}[1]{>{\centering\arraybackslash}p{#1}}
\newcolumntype{M}[1]{>{\centering\arraybackslash}m{#1}}
\usepackage{url,hyperref}
\usepackage[usenames,dvipsnames]{xcolor}
\hypersetup{colorlinks=true, linkcolor=BrickRed, urlcolor=blue!50!black, citecolor=blue!50!black}
\begin{document}
 
\title{Knots are Generic Stable Phases in Semiflexible Polymers}
\author{Suman Majumder}\email[]{suman.majumder@itp.uni-leipzig.de}
\affiliation{Institut f\"ur Theoretische Physik, Universit\"at Leipzig, IPF 231101,
04081 Leipzig, Germany}

\author{Martin Marenz}\email[]{martin.marenz@itp.uni-leipzig.de}
\affiliation{Institut f\"ur Theoretische Physik, Universit\"at Leipzig, IPF 231101,
04081 Leipzig, Germany}
\author{Subhajit Paul}\email[]{subhajit.paul@itp.uni-leipzig.de}
\affiliation{Institut f\"ur Theoretische Physik, Universit\"at Leipzig, IPF 231101,
04081 Leipzig, Germany}
\date{\today}
\author{Wolfhard Janke}\email[]{wolfhard.janke@itp.uni-leipzig.de}
\affiliation{Institut f\"ur Theoretische Physik, Universit\"at Leipzig, IPF 231101,
04081 Leipzig, Germany}
\date{\today}
 
\begin{abstract}

  \begin{center}
   \textbf{Abstract}
  \end{center}
Semiflexible polymer models are widely used as a paradigm to understand structural phases in 
  biomolecules including folding of proteins. Since stable knots are not so common in real proteins, the existence of stable knots in semiflexible polymers has not been explored much. Here, via extensive replica exchange Monte Carlo simulation we investigate the same for a bead-stick and a bead-spring homopolymer model that covers the whole range from flexible to stiff. We establish the fact that the presence of stable knotted phases in the phase diagram is dependent on the ratio $r_b/r_{\rm{min}}$ where $r_b$ is the equilibrium bond length and  $r_{\rm{min}}$ is the distance for the strongest nonbonded contacts. Our results provide evidence for both models that if the ratio $r_b/r_{\rm{min}}$ is outside a small window around unity then depending on the bending stiffness one always encounters stable knotted phases along with the usual frozen and bent-like structures at low temperatures. These findings prompt us to conclude that knots are generic stable phases in semiflexible polymers.   
\end{abstract}
 
\maketitle
\section{Introduction}
Identification and prediction of macromolecular conformations via computer simulations have developed so much 
over the years with current possibility of doing them at atomistic or even at quantum level. Those detail simulations are always limited by their accessible time scales which often may be too small to provide meaningful insight or it may require enormous effort to arrive at the same. On the other hand, computationally less expensive coarse-grained approaches (by integrating out certain unimportant degrees of freedom) are more than sufficient to understand the generic behavior of macromolecules \cite{muller2002coarse}. 
The very simple idea of consideration of self-avoidance and introduction of attraction led to the exploration 
of theta polymers that are characterized by collapse and freezing transitions \cite{Lifshitz1978}. In this regard, even 
simplest lattice models could provide a deep insight \cite{carmesin1988bond,grassberger1995simulations,bastolla1997phase,rampf2005first,vogel2007freezing,farris2018role}. 
The more generic off-lattice models come in two major variants, viz., the bead-stick model and the bead-spring model. They have 
widely been used to investigate different structural phases of polymers, be it a single one 
\cite{schnabel2009elastic,schnabel2009surface,seaton2010collapse,seaton2013flexible,marenz2016knots,zierenberg2016dilute} or in aggregates \cite{zierenberg2015amorphous,ranganathan2016,zierenberg2016dilute,midya2019phase} 
in bulk, and in some cases on surfaces \cite{sintes2001adsorption,moddel2014adsorption,austin2017interplay,oberthur2018two,milchev2019linear} or under geometrical constraints \cite{zierenberg2014aggregation,milchev2018densely}.
\par
While dealing with these models one should always be aware of the effect of bending stiffness that is used 
as a parameter to distinguish a flexible polymer from a semiflexible or a stiff one. In this context, a simple worm-like chain 
model is sufficient to emulate bending-energy-dominated polymers or semiflexible polymers \cite{kratky1949}. Such an approach nicely mimics several features of complex biopolymers that includes DNA, RNA, and even some proteins. Since the worm-like chain model does not take the self-avoidance or any nonbonded interactions into consideration, it fails to capture the structural transitions associated with a theta polymer. Thus for a complete understanding combining the features of theta polymers with the bending stiffness is necessary. 
From this point of view, using a 
bead-spring model, Seaton \textit{et al.}\ \cite{seaton2013flexible} explored different phases (coiled, collapsed, frozen, bent, hairpin and toroidal conformations) 
of a semiflexible polymer just by tuning the bending stiffness. Recently we have shown that similar phases can 
also be realized if a bead-stick model is used instead \cite{marenz2016knots}. Intriguingly, in addition to those phases there 
we have discovered new pseudo phases characterized by thermodynamically stable knotted structures of the polymer. 
\par
Strictly, knots are topological properties of closed strings, and hence, knots found in open polymers are 
not mathematically defined \cite{kauffman1991}. Nevertheless, by means of a special strategy for ring closure, the definition can be extended to open 
polymers as well. Thus, the presence of knots in polymers has fascinated chemists and physicists for long 
\cite{frisch1961chemical,frank1975statistical,liu1976knotted,koniaris1991knottedness,taylor2003protein,virnau2005knots}. 
Especial interest has evolved around investigating knots in proteins which are best described by semiflexible polymer models. Those studies reveal that only a small fraction of them form knots \cite{taylor2000deeply,lua2006statistics,virnau2006intricate,jamroz2014knotprot}. There have been attempts to understand this fact by arguing that knotted proteins are evolutionary 
unfavorable \cite{wust2015sequence}. 
\par
In contrast to proteins, the chances of 
realizing a knot are higher in flexible polymers either in the swollen or globular phase 
\cite{koniaris1991knottedness,deguchi1997universality,lua2004fractal,virnau2005knots}. The knots identified in most of these studies 
are formed by chance and cannot be considered to characterize true thermodynamically stable phases. Only recently, in our 
simulations of a bead-stick semiflexible polymer model where almost the whole range of possible bending stiffnesses was explored, 
we found stable knots \cite{marenz2016knots}. However, as mentioned earlier in their comprehensive study 
of the phase diagram of a semiflexible polymer using a bead-spring model, Seaton \textit{et al.}\ \cite{seaton2013flexible} did not mention 
any presence of knotted conformations. This poses the important question whether knots 
are generic phases only in bead-stick polymers. In Ref.\ \cite{marenz2016knots} it has been conjectured that 
the formation of knots is dependent on the ratio of the equilibrium bond length $r_b$ and the distance $r_{\rm{min}}$ for which the energy due to nonbonded contacts attains its minimum. In this work we take up this task and study how the ratio $r_b/r_{\rm{min}}$ influences the presence 
of stable knot phases in the phase diagram using both a bead-stick and a bead-spring model. Using the bead-spring model 
will be particularly helpful in explaining the missing knots in the model used in Ref.\ \cite{seaton2013flexible}. Our results provide evidence that for both the 
bead-stick and the bead-spring model knotted structures form a stable phase covering a range 
of bending stiffnesses if $r_b/r_{\rm{min}}$ is away from a small region around unity. This can be explained by analyzing the competition between the nonbonded energy minimization and the bending energy minimization.
\par
The rest of the paper is organized as follows. Next in the Sec.\ \ref{model} we explain the two different models we will be using, the setup of the replica exchange simulation method and the data analysis procedure. The details of the bead-spring model employed in Ref.\ \cite{seaton2013flexible} are relegated to the Appendix. The results are presented in Sec.\ \ref{result}. Finally, we put forward our conclusions in Sec.\ \ref{conclusion}. 

\section{Simulation Details}\label{model}
\subsection{Models}
As already outlined above we consider two semiflexible polymer models: (i) bead-stick and (ii) bead-spring. In both models 
the monomers are considered to be spherical beads with diameter $\sigma$, and the nonbonded interaction energy 
is dependent on the inter-particle distance $r_{ij}$ and is given as
\begin{eqnarray}\label{potential_OLM}
E_{\rm {nb}}=\sum_{i=1}^{N-2}\sum_{j=i+2}^{N} \left[ E_{\rm {LJ}}({\rm{min}}\{r_{ij},r_c\})-E_{\rm {LJ}}(r_c) \right] 
\end{eqnarray}
where 
\begin{eqnarray}\label{our_LJ}
E_{\rm {LJ}}(r_{ij})=4\epsilon \left[ \left( \frac{\sigma}{r_{ij}} \right)^{12} - \left (\frac{ \sigma}{r_{ij}} \right )^{6} \right]
\end{eqnarray}
is the standard Lennard-Jones (LJ) potential which has a minimum at $r_{\rm{min}}=2^{1/6}\sigma$. In Eq.\ \eqref{potential_OLM}, $N$ is the length of the polymer measured as the total number of beads or monomers. In order to be consistent with our previous study \cite{marenz2016knots} for the bead-stick model we set $\sigma=1.0$ and do not use any cut-off in $E_{\rm{nb}}$, whereas for the bead-spring model we choose $\sigma=2^{-1/6}$ in order to be consistent with the choice of $r_{\rm{min}}=1.0$ in Ref.\ \cite{seaton2013flexible}  and set $r_c=2.5\sigma$ for faster computation of $E_{\rm {nb}}$. For both models the nonbonded interaction 
strength $\epsilon$ is set to unity. In bead-stick models the monomers form a chain where the connectivity 
between successive monomers are maintained via rigid bonds having fixed length $r_b$. On the other hand, in a bead-spring model the bonds between successive monomers 
are maintained via some kind of springs. Here we consider the standard 
finitely extensible non-linear elastic (FENE) potential \cite{milchev1993off,milchev2001formation}
\begin{eqnarray}\label{FENE}
E_{\rm{FENE}}=-\frac{K}{2}R^2 \sum_{i=1}^{N-1}\ln\left[1-\left(\frac{r_{ii+1}-r_b}{R}\right)^2\right]
\end{eqnarray}
where $r_b$ is the equilibrium bond distance for which $E_{\rm{FENE}}$ is minimum. Unless otherwise mentioned in all the simulations we have used $R=0.3$ and $K=40$. 
\par
In both models stiffness is introduced via the well-known 
discretized worm-like chain cosine potential given as 
\begin{equation}\label{stiff}
 E_{\rm{bend}}=\kappa \sum_{i=1}^{N-2}(1-\cos \theta_i)
\end{equation}
where $\theta_i$ is the angle between consecutive bonds and $\kappa$ controls the effective bending stiffness of the polymer. In this work we aim to perform simulations of the two models using different values of $r_b/r_{\rm{min}}$. For that we fix the value 
of $r_{\rm{min}}=2^{1/6}$ and $1.0$, respectively, for the bead-stick and the bead-spring model 
(by keeping the respective values of $\sigma$ in all our simulations) and vary only the equilibrium bond length $r_b$.

\subsection{Simulation method}
It is known that the phase diagram of coarse-grained semiflexible polymers contains ``strong'' first-order phase transitions, where ``strong''
means that the two coexisting phases are separated in phase space by a highly suppressed
region \cite{noguchi1997first,marenz2016knots}. On top of that, such systems obey very slow dynamics at low temperatures, even far away
from these phase transitions. This demands application of relatively complex Monte
Carlo (MC) simulation methods to obtain well equilibrated results \cite{marenz2016knots,janke2018macromolecule}. Previously we have used a 
parallelized version of the multicanonical algorithm \cite{berg1991multicanonical,zierenberg2013scaling,janke2016SM} along with replica exchange (RE) (also known as parallel tempering) \cite{hukushima1996exchange} and the two-dimensional replica exchange method (2D-RE) \cite{marenz2016knots}. Both of them were shown to produce the same 
results and hence here, we restrict ourselves to use only the 2D-RE algorithm. It is based on many individual Metropolis MC
simulations which run in parallel, each at a different parameter pair ($T,\kappa$), whose conformations are exchanged every now and then. This substantially improves the quality of the 
canonical estimates near the phase transitions and also at low temperatures. 
\par
For 2D-RE it is necessary to write down the system Hamiltonian in the following form
\begin{equation}\label{hamiltonian}
 H=E_0+\kappa E_1
\end{equation}
where $E_0$ corresponds to the base energy coming from the nonbonded interaction defined in Eq.\ \eqref{potential_OLM} and the bonded 
interaction (if any) in Eq.\ \eqref{FENE}, and $E_1$ corresponds to the energy contribution coming from the bending stiffness term $\sum_{i=1}^{N-2}(1-\cos \theta_i)$ defined 
in Eq.\ \eqref{stiff}. While interchanging replicas between two neighboring points $\mu$ and $\nu$, in the simulation parameter space ($T, \kappa$), the above 
splitting of the total energy is used to calculate the exchange probability as 
\begin{equation}\label{exchange_p}
 p(\mu \leftrightarrow \nu) =\min \left[1, \exp(\Delta \beta \Delta E_0+ \Delta (\beta \kappa) \Delta E_1 ) \right],
\end{equation}
where $\beta=1/k_{\rm B}T$ ($k_{\rm B}=1$ being the Boltzmann constant).
The two-dimensional parameter space has the advantage that it can avoid topological barriers which would
hinder the flux in a one-dimensional parallel tempering simulation. In one-dimensional parallel
tempering simulation it can happen that there are some temperatures $T$ where
almost no state exchange occurs which can be avoided in 2D-RE via exchange along the other direction ($\kappa$) in the parameter space $(T, \kappa)$.

\par
Apart from the 2D-RE algorithm, it is also necessary to adapt different MC update moves to tackle the underlying problem. The set of updates includes 
the usual crank-shaft, spherical-rotation, and pivot moves for both the bead-stick and the bead-spring models \cite{Austin2018}. For the bead-spring model we have also used 
the single monomer displacement moves. In addition to these standard but simple moves we have also used the complex double-bridge and bridge-end moves \cite{pant1995variable,karayiannis2002novel}. Note that for the bead-stick model the bridge moves are adjusted accordingly to respect the 
fixed bond lengths.

\subsection{Analysis} Pursuing the 2D-RE simulations allows us to use the two-dimensional version of the weighted histogram analysis 
method (2D-WHAM) for generating appropriate canonical estimates of quantities of interest \cite{ferrenberg1988new,kumar1992weighted}. 
Here, one starts by measuring two-dimensional histograms $H_i(E_0,E_1)$ at $m$ different parameter pairs $(T,\kappa)_i$ which gives the energy 
distribution 
\begin{equation}\label{2d-hist}
p_i=\frac{H_i(E_0,E_1)}{N_i} 
\end{equation}
where $N_i$ is the number of measurements done at each individual parameter space-point $(T,\kappa)_i$ to generate $H_i(E_0,E_1)$. 
Using this one writes down the density of states as 
\begin{equation}\label{dos}
\Omega(E_0,E_1)=\frac{\sum_{k=1}^{m}{g_k}^{-1}p_k(E_0,E_1)}{\sum_{k=1}^{m} N_k {g_k}^{-1} Z_{\beta_k,\kappa_k}^{-1}\exp[-\beta_k(E_0+\kappa E_1)]},
\end{equation}
where ${g_k}=1+2\tau_k$ accounts for the integrated autocorrelation time $\tau_k$ calculated via 
binning analysis from the time series generated at each parameter point $k$. 
In Eq.\ \eqref{dos}, the partition function is given as 
\begin{equation}\label{partition}
 Z_{\beta_i,\kappa_i}=  \sum_{E_0,E_1}
 \frac{\sum_{k=1}^{m}{g_k}^{-1}p_k(E_0,E_1)}{\sum_{k=1}^{m} N_k {g_k}^{-1} Z_{\beta_k,\kappa_k}^{-1}\exp[-\beta_k(E_0+\kappa E_1)]}\exp[-\beta_i(E_0+\kappa E_1)].
\end{equation}
Note that \textit{a priori} neither $\Omega(E_0,E_1)$ nor $Z_{\beta_i,\kappa_i}$ is known. Assuming appropriate initial values of $Z_{\beta_i,\kappa_i}$, the self-consistent Eqs.\ \eqref{dos} and \eqref{partition} are solved to arrive at precise values of $\Omega(E_0,E_1)$ and $Z_{\beta_i,\kappa_i}$ \cite{janke2013monte}. Once this is done 
the estimate of any observable $O$ at any parameter point $(T,\kappa)$ can be calculated via
\begin{equation}\label{estimate}
 \langle O \rangle_{\beta,\kappa}=\frac{\sum_{E_0,E_1} O(E_0,E_1) \Omega(E_0,E_1)\exp[-\beta(E_0+\kappa E_1)]}{\sum_{E_0,E_1}\Omega(E_0,E_1)\exp[-\beta(E_0+\kappa E_1)]}.
\end{equation}
In our analyses, we will be dealing with derivatives of certain observables with 
respect to temperature $T$ calculated as 
\begin{equation}\label{dT}
 \frac{d }{dT}\langle O \rangle=k_{\rm B}\beta^2 \left(\langle OE \rangle -\langle O \rangle \langle E \rangle \right)
\end{equation}
where $E=E_0+\kappa E_1$ is the total energy. The statistical errors on all the observables are obtained via the Jackknife method \cite{efron1982jackknife}.
Different observables that we measure during our simulations will be explained subsequently in the results section.
\begin{figure*}[t!]
\includegraphics[width=0.45 \textwidth]{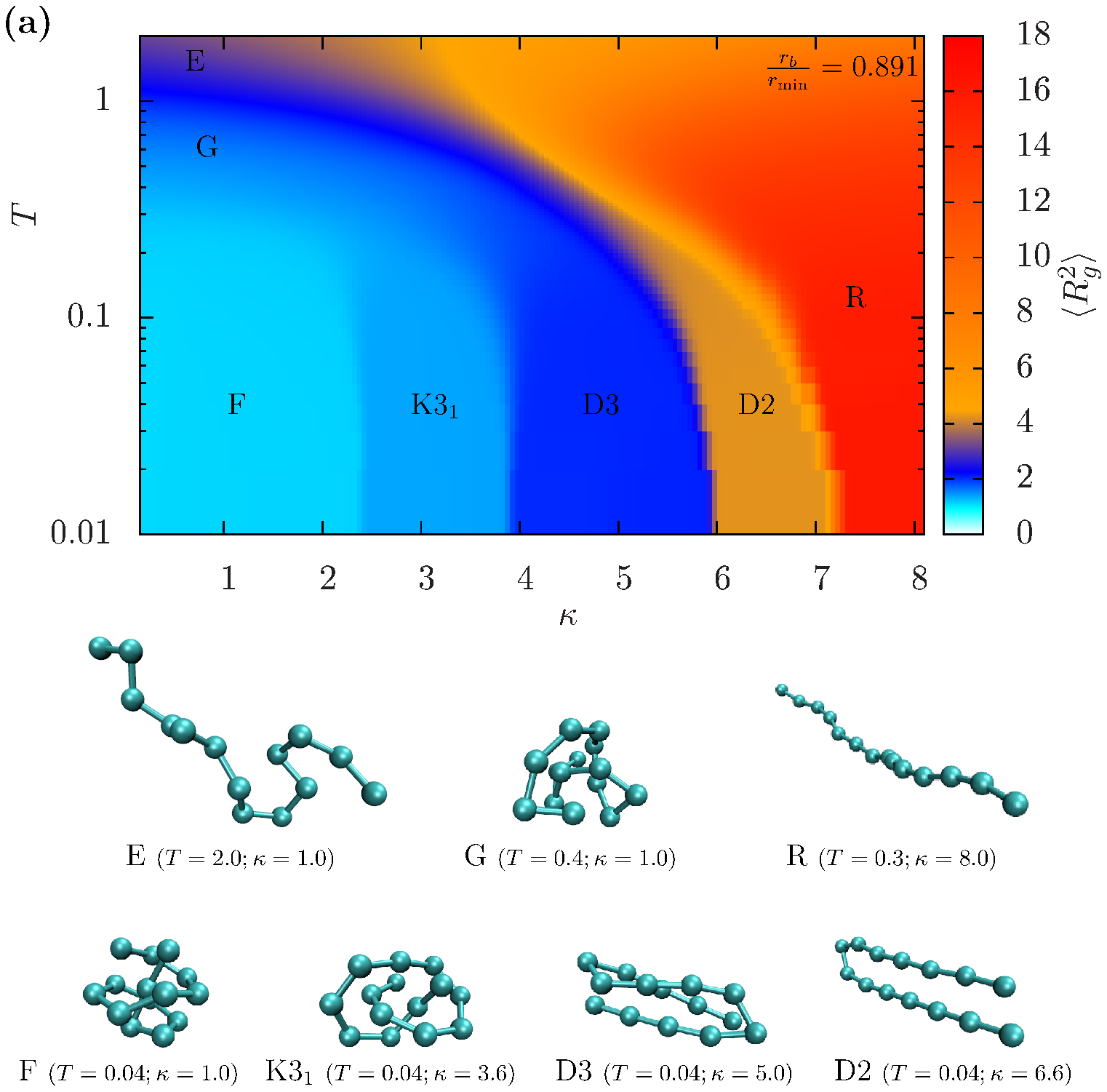}
~~\includegraphics[width=0.45 \textwidth]{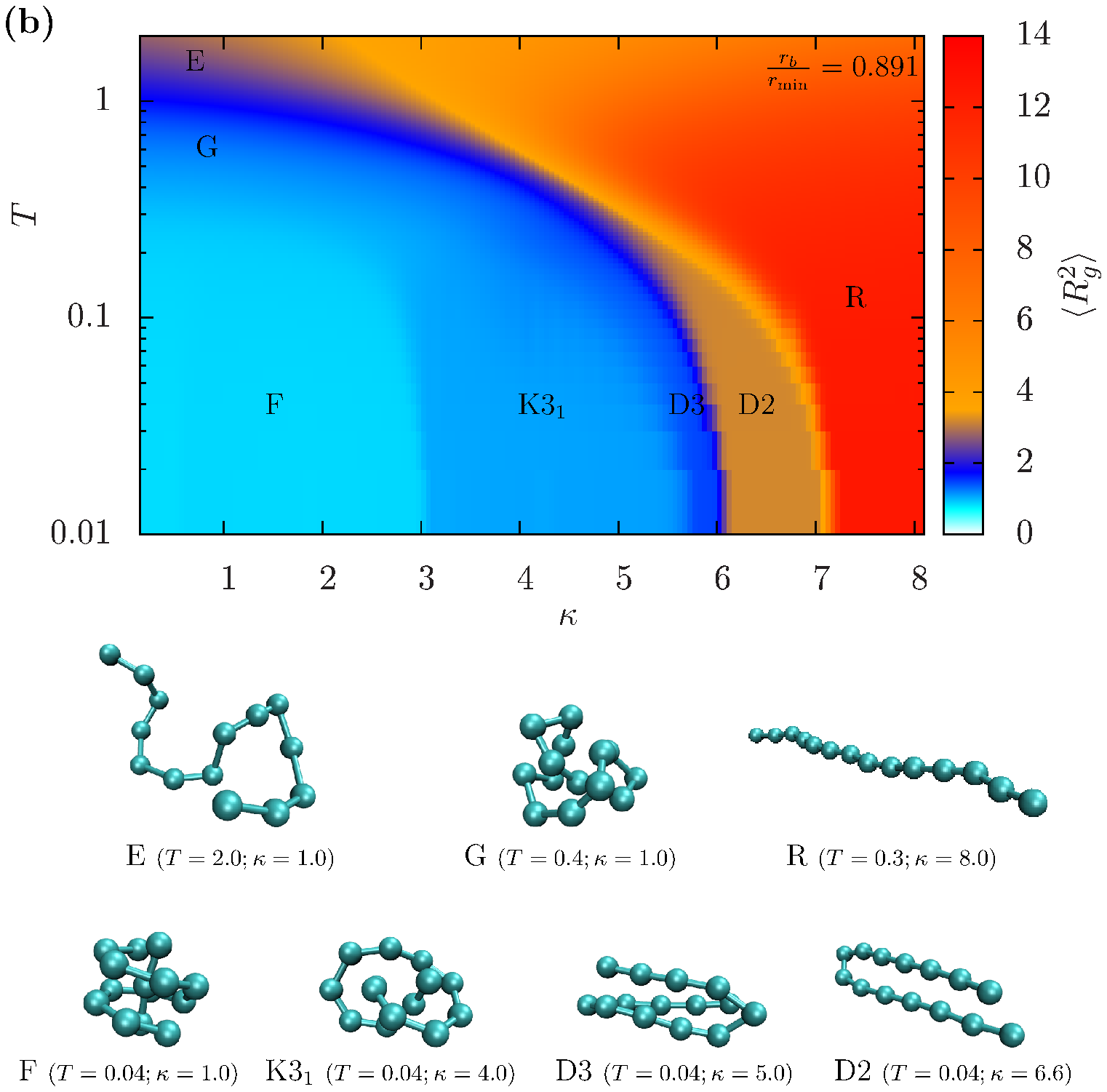}
\caption{Complete phase diagram for (a) the bead-stick model and (b) the bead-spring model with the usual choice of the ratio $r_b/r_{\rm{min}}=0.891$ for $N=14$. The surface plots are drawn with respect to the spatial extension of the polymer chain measured in terms of the squared radius of gyration $\langle R_g^2\rangle$. The labeled phases stand for the following:
E for elongated; R for rod-like; G for globular; F for frozen; $\rm{K}C_n$ for knotted phase with the corresponding knot type $C_n$; D$n$ for bent phases with $n$ number of segments.}
\label{Full-PD}
\end{figure*}
\section{Results}\label{result}
As already mentioned we aim to explore the effect of the ratio $r_b/r_{\rm{min}}$ on the presence of stable knotted phases or in general different phases in both models described above. Thus subsequently all the results are organized with respect to the choice of $r_b/r_{\rm{min}}$. In the following we report results for polymers of length $N=14$ and $28$. This choice is motivated by the not too high complexity of the pseudo-phase diagrams and at the same time will be sufficient to understand the effect of varying $r_b/r_{\rm{min}}$ on the existence of stable knots.
\begin{figure*}[t!]
\includegraphics[width=0.4 \textwidth]{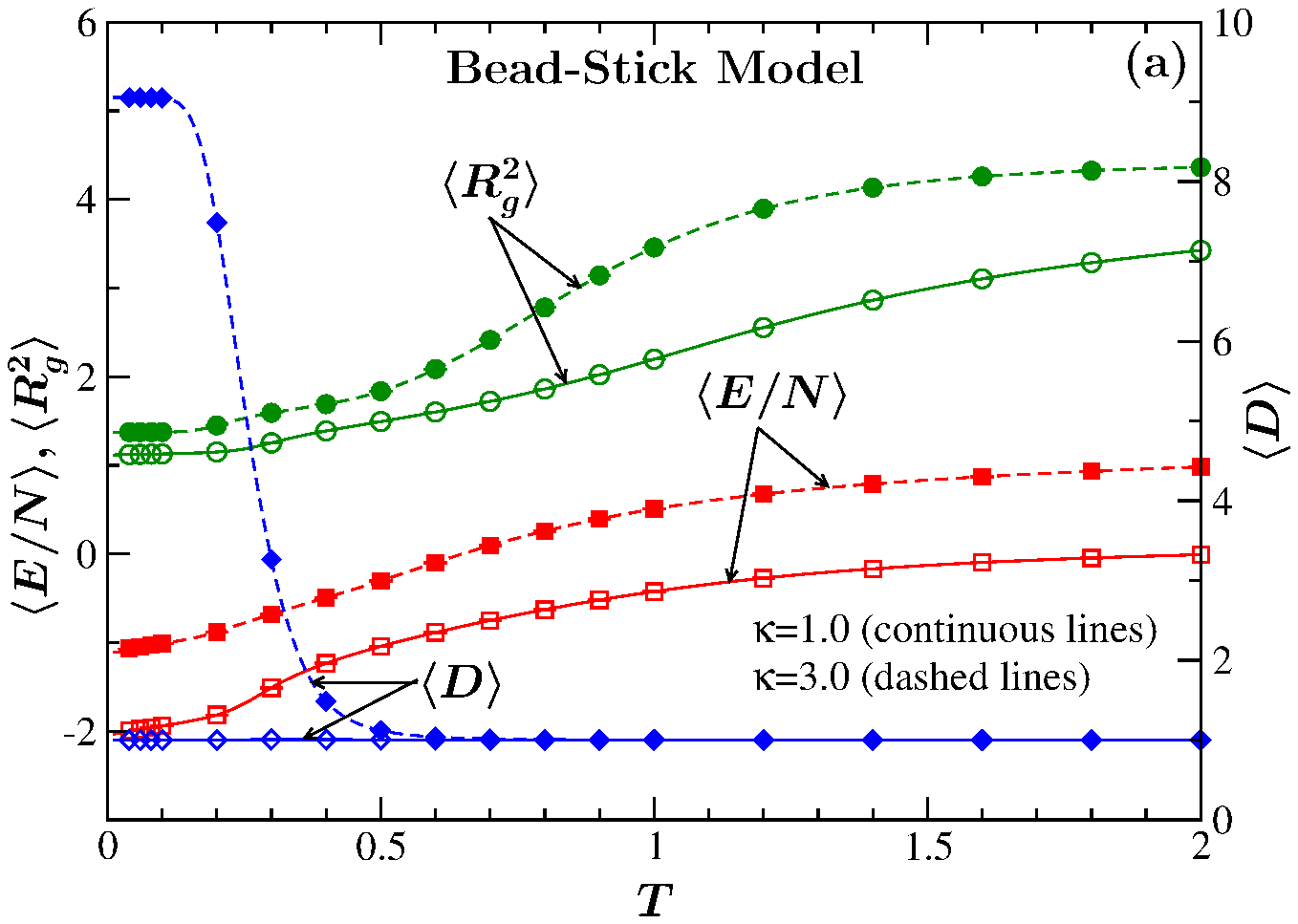}
~~\includegraphics[width=0.4 \textwidth]{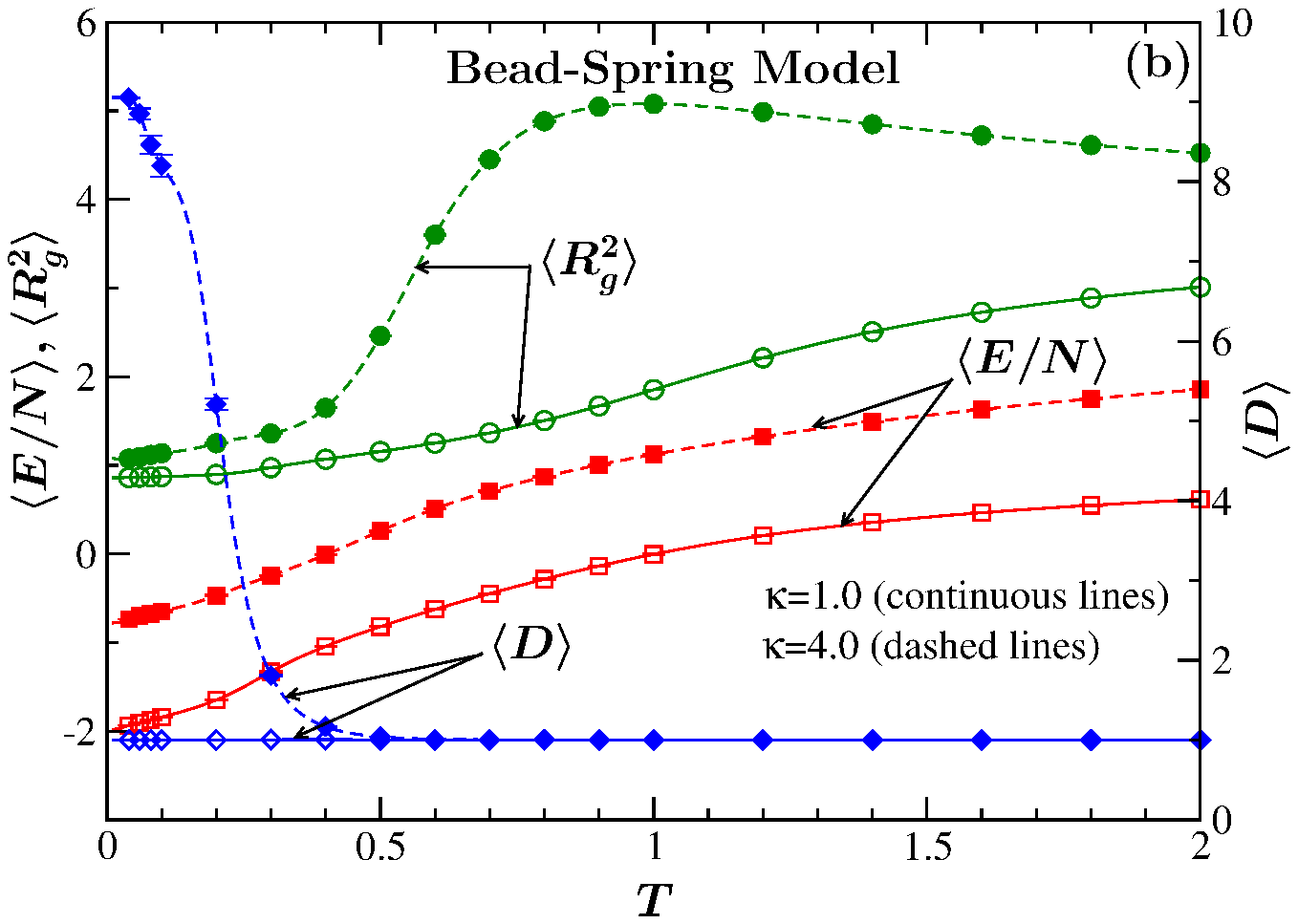}\\
\vskip 0.05cm
~~\includegraphics[width=0.41 \textwidth]{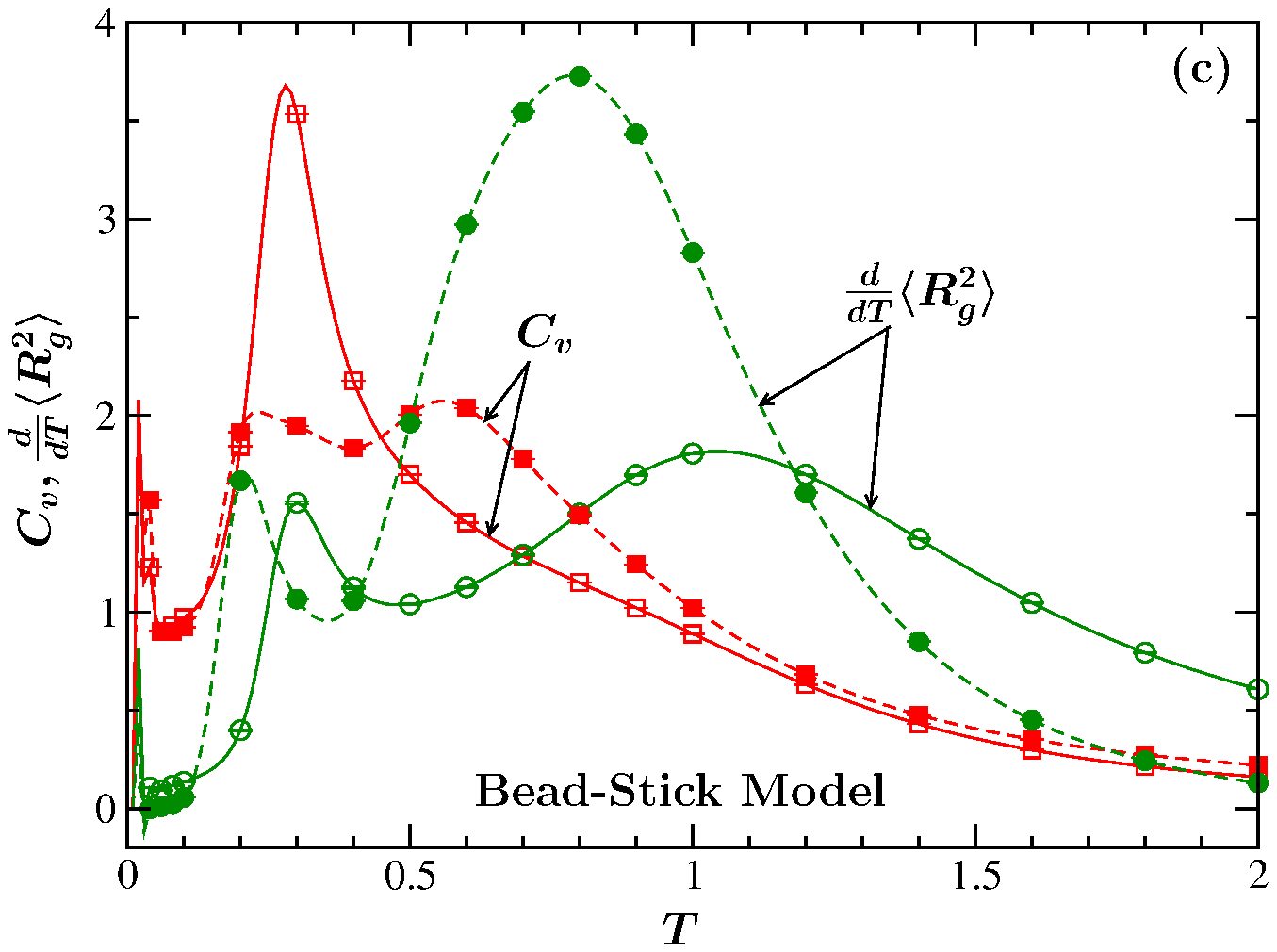}
\includegraphics[width=0.42 \textwidth]{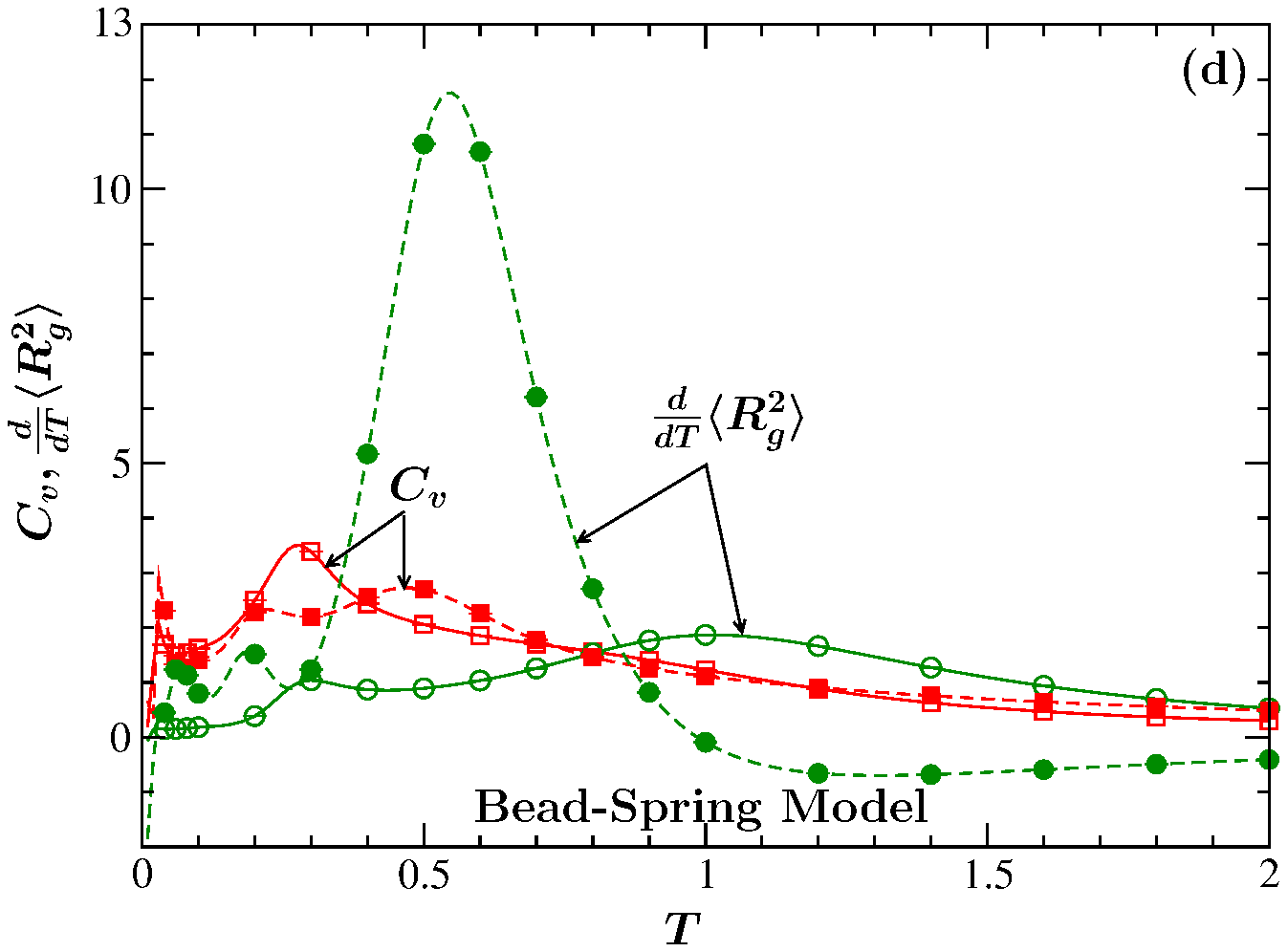}
  \caption{Plots in (a) and (b) show the validity of the measured quantities, viz., energy density $\langle E/N\rangle$, the squared radius of gyration $\langle R_g^2\rangle$, and the knot parameter $\langle D \rangle$, to identify the expected transitions between different phases, respectively for the two models. The quantities are plotted as a function of temperature $T$ with two different choices of the bending stiffness $\kappa$ as mentioned within (a) and (b). Plots in (c) and (d) show the corresponding plots for indirectly measured quantities, viz., specific heat $C_v={d \langle E\rangle}/{dT}$ and the derivative of the squared radius of gyration ${d\langle R_g^2 \rangle}/{dT}$, for the two models. All the data presented here are for the choice of $r_b/r_{\rm{min}}=0.891$ and $N=14$.}
  \label{parameters}
\end{figure*}
\subsection{Phase behavior for $r_b/r_{\rm{min}}=0.891$}
We start our investigation with the choice of $r_b/r_{\rm{min}}=2^{-1/6}\approx0.891$, as was used for the bead-stick model in Ref.\ \cite{marenz2016knots}. Figures\ \ref{Full-PD}(a) and (b) show the complete phase diagram in the temperature $T$ and bending stiffness $\kappa$ plane, for both models with a chain length $N=14$. The surface plot to differentiate between the different phases is obtained by using the estimated squared radius of gyration $\langle R_g^2\rangle$ calculated as 
\begin{equation}
 R_g^2=\frac{1}{2N^2}\sum_{i,j=1}^N(\vec{r_i}-\vec{r_j})^2
\end{equation}
where $\vec{r_i}$ is the position vector of the $i$-th monomer. $R_g^2$ gives a measure of the spatial extension of the polymer. For both models, a rich variety of phases can be observed. Elongated (E) and rod-like (R) conformations are obtained as the two major structures in the noncondensed state, respectively at low and high $\kappa$. In the condensed phases, depending on the bending stiffness and temperature one observes interesting conformations that range from usual frozen state (F) to bent phases (D$n$). Most importantly, like in the bead-stick model (already demonstrated in Ref.\ \cite{marenz2016knots}), the bead-spring model, too, shows the existence of a knotted phase in the range $\kappa \in [3.2,5.8]$ which is even wider than the corresponding range $\kappa \in [2.6,3.8]$ for the bead-stick model. 

\begin{figure*}[b!]
\includegraphics[width=0.4 \textwidth]{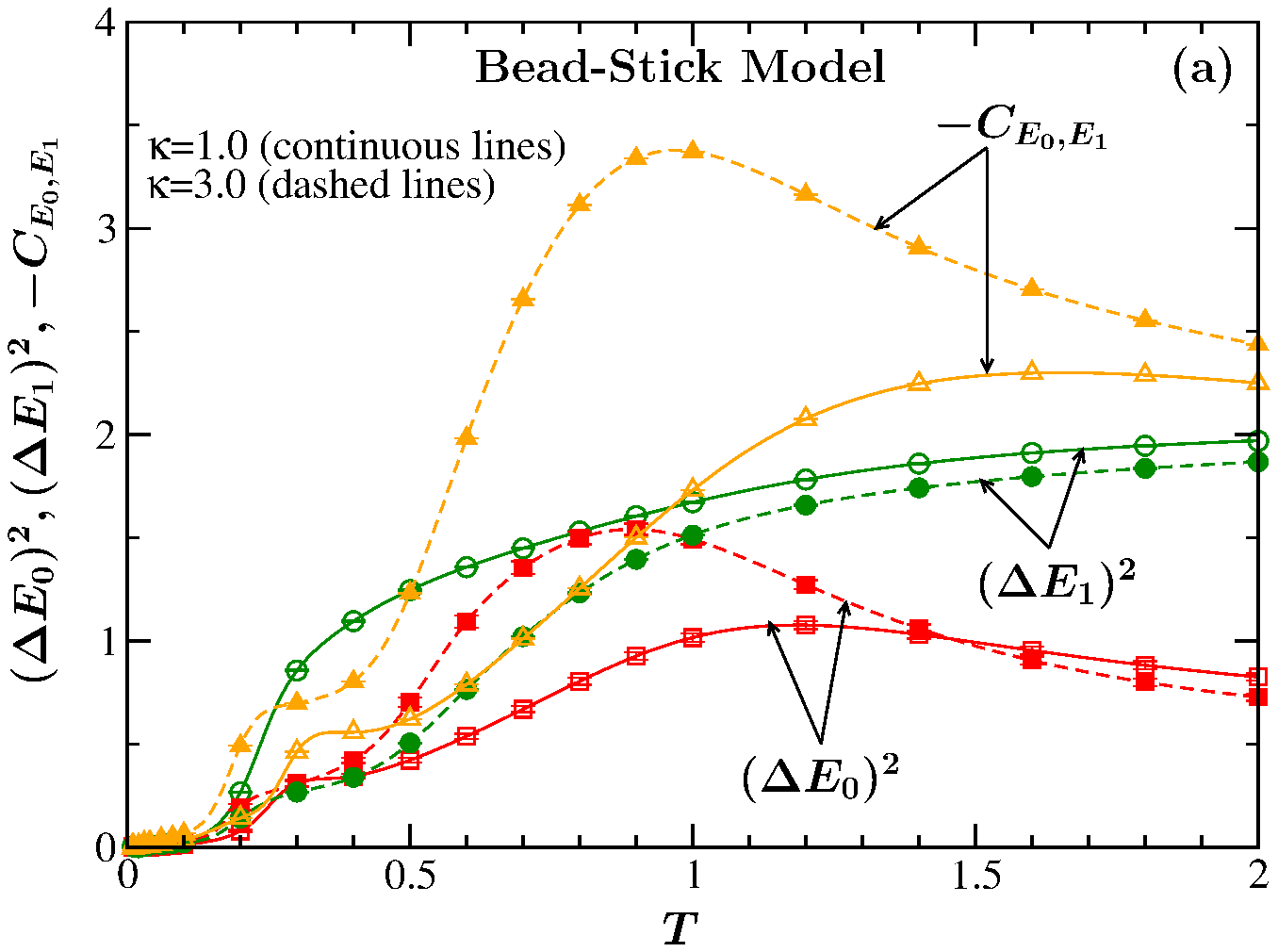}
~~~~~~\includegraphics[width=0.4 \textwidth]{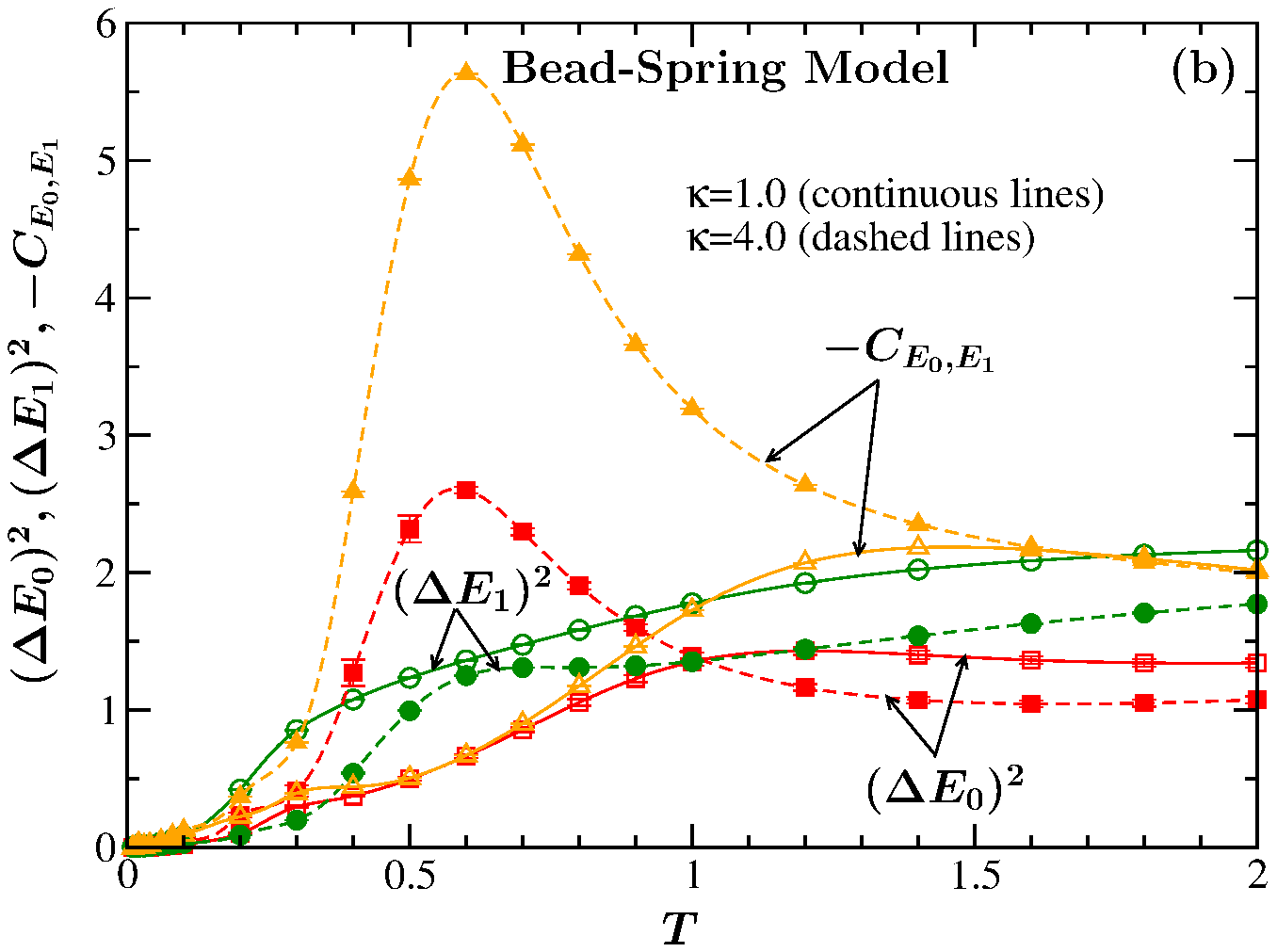}
  \caption{Variance of the base energy $(\Delta E_0)^2$ and the bending energy $(\Delta E_1)^2$, and their cross-correlation $C_{E_0,E_1}$ as a function of temperature with two different choices of $\kappa$ for (a) the bead-stick and (b) the bead-spring model. As in Fig.\ \ref{parameters}, the data are for the choice of $r_b/r_{\rm{min}}=0.891$ and $N=14$.}
  \label{fluctuation}
\end{figure*}

\par
Before we proceed further with other values of the $r_b/r_{\rm{min}}$ parameter it would be worth to limit 
ourselves to the quantities which are relevant for identifying the knotted phase. For that in the present case following the custom we have estimated from our simulation data the energy density $\langle E/N\rangle$ along with $\langle R_g^2 \rangle$. Both these quantities for a fixed $\kappa$ do not show any signature of pseudo-phase transition (strictly the term phase transition is used in the thermodynamic limit, i.e., in the large $N$ limit), as evident from the corresponding plots for both the bead-stick and bead-spring model presented, respectively in Figs.\ \ref{parameters}(a) and (b). The cases for the higher 
value of $\kappa=3.0$ and $4.0$ (shown by the dashed lines in the figure), respectively, for the two models correspond to values within the knotted phase. In fact, these parameters also do not provide a strong evidence even for freezing or collapse transition as expected for the lower $\kappa=1.0$ [shown by the continuous lines in Figs.\ \ref{parameters}(a) and (b)] for both models. For this matter, one can also look at the corresponding derivatives using Eq.\ \eqref{dT}, i.e., the specific heat $C_v=\frac{d\langle E \rangle}{dT}$ and $ \frac{d\langle R_g^2 \rangle}{dT}$ 
which are presented for both the $\kappa$ values in Figs.\ \ref{parameters}(c) and (d), respectively for the two models. The derivative $\frac{d\langle R_g^2 \rangle}{dT}$ seems to provide a clear signature for the collapse transition for both models. For the bead-stick model the collapse transition temperatures for $\kappa=1.0$ and $3.0$ can roughly be read off as $0.85$ and $1.1$, respectively which can also be appreciated with regards to the phase diagram presented in Fig.\ \ref{Full-PD}(a). Similarly, in case of the bead-spring model, the data for $ \frac{d\langle R_g^2 \rangle}{dT}$ provides a reasonable signature of the collapse transition temperatures for both the $\kappa$ values. The specific heat $C_v$ for both models show peaks at some respective temperatures that may be identified as the collapse transition temperature. However, they are located at values substantially lower than the 
corresponding values obtained from $\frac{d\langle R_g^2 \rangle}{dT}$. The 
low-temperature peaks for the  $\frac{d\langle R_g^2 \rangle}{dT}$ data are prominent for the lower $\kappa$ values for both models which correspond to the transition to the frozen state $F$. On the other hand, at low temperature, peaks for the higher $\kappa$ values for both models are not so pronounced to mark the transition to the stable knotted phase. 
\begin{table}[b!]
\caption{Expressions for Alexander polynomial $\Delta(t)$ and the corresponding unique knot parameter $D=\Delta_p(-1.1)$ for some simple knots which we encounter in this work.}\label{tab1_5}
\centering
\renewcommand{\arraystretch}{2}
\begin{tabular}{{|P{1.3cm}|P{2.0cm}|P{3.0cm}|P{1.3cm}|}}

\hline

&schematic&Alexander polynomial $\Delta(t)$&$\Delta_p(-1.1)$\\
\hline
\hline

unknotted&\includegraphics[width=0.1 \textwidth]{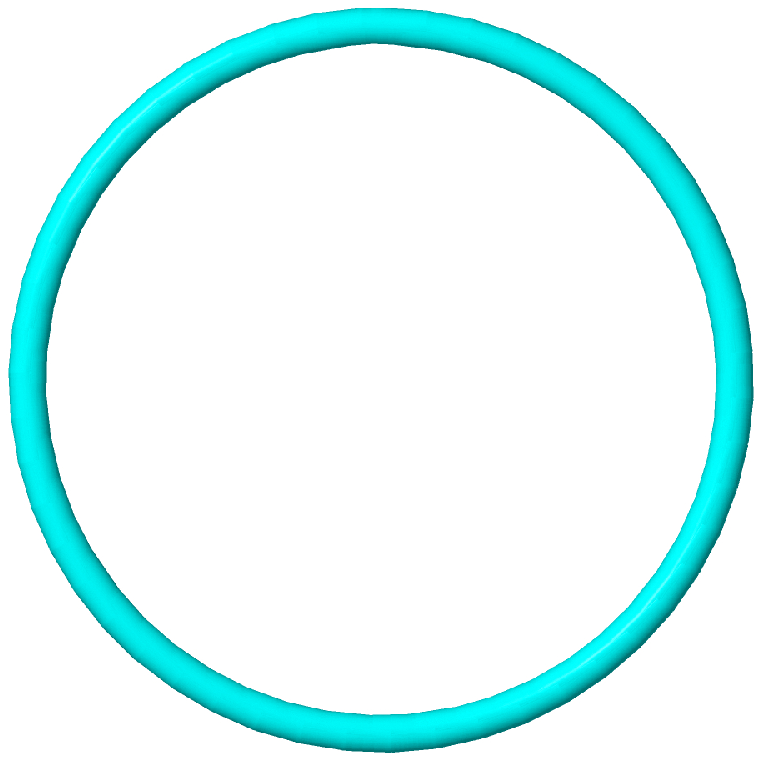}&1&$1.0$\\

$3_1$&\includegraphics[width=0.1 \textwidth]{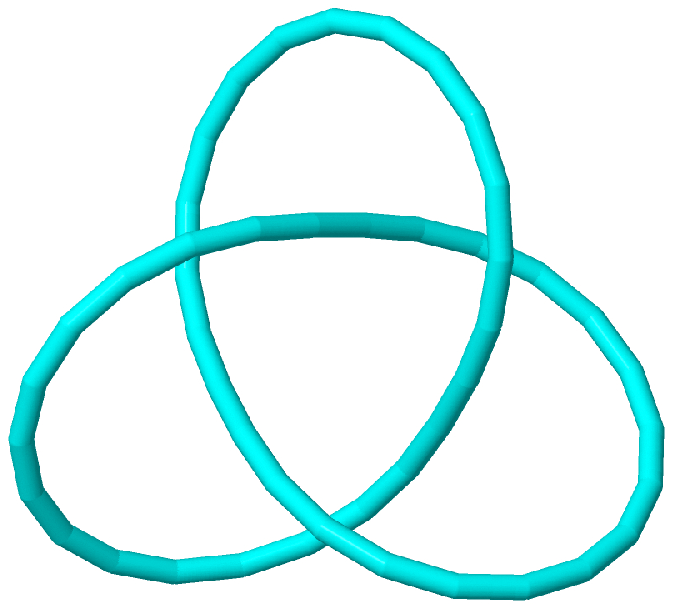}&$t+t^{-1}-1$&$9.05462$\\
$4_1$&\includegraphics[width=0.1 \textwidth]{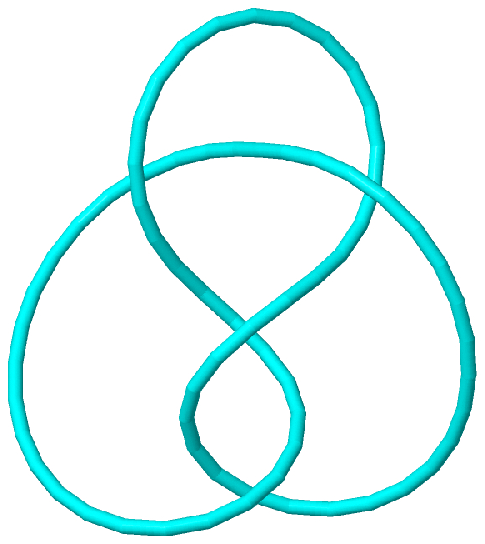}&$-t-t^{-1}+3$&$25.09099$\\
$5_1$&\includegraphics[width=0.1 \textwidth]{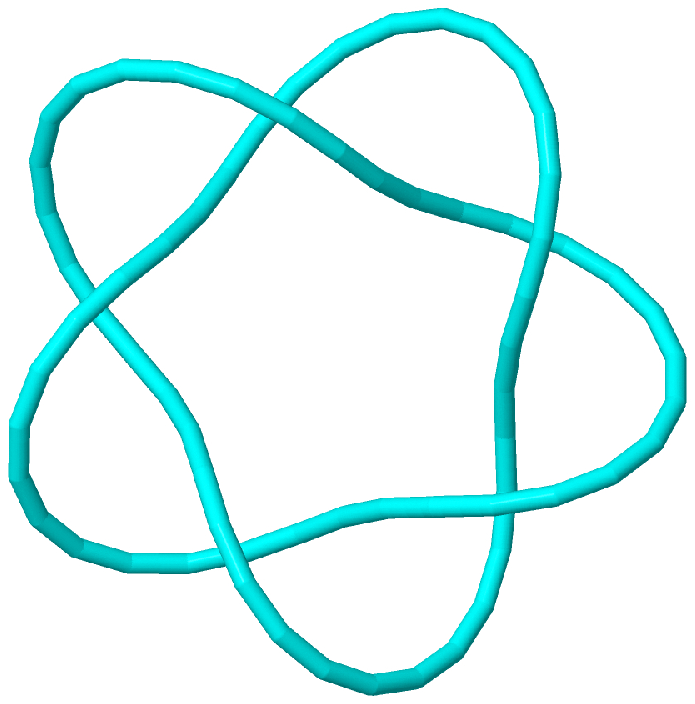}&$t^2+t^{-2}-t-t^{-1}+1$&$25.45745$\\
$8_{19}$~&\includegraphics[width=0.1 \textwidth]{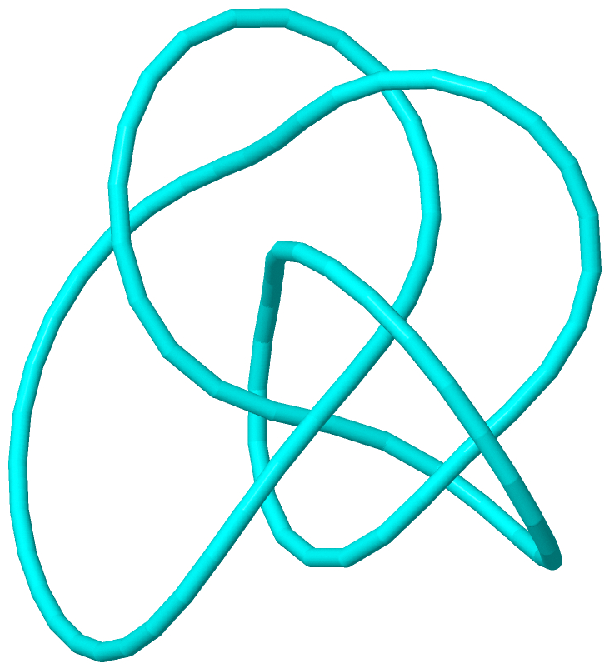}&$t^3+t^{-3}-t^2-t^{-2}+1$&
$9.72667$\\
\hline
\hline
\end{tabular}

\end{table}
\par
{In Ref.\ \cite{marenz2016knots} using the bead-stick model it has been pointed out that the transition K$3_1$$\leftrightarrow$D$3$ is first order which is signaled by a bimodal distribution in the two-dimensional space of energies $E_0$ and $E_1$. In view of that we estimate the variances
\begin{equation}
 (\Delta E_0)^2=\langle E_0^2 \rangle -\langle E_0\rangle^2
\end{equation}
and 
\begin{equation}
 (\Delta E_1)^2=\langle E_1^2 \rangle -\langle E_1\rangle^2,
\end{equation}
respectively, for the base energy and the bending energy separately. In Fig.\ \ref{fluctuation} the corresponding plots are shown as a function of temperature with the same choices of $\kappa$ as in Fig.\ \ref{parameters}, for both the models. Clearly, the data do not provide any significant signature of the transition to a knotted phase. As it is intuitive that different phases in a semiflexible polymer result from the interplay of the base energy and the bending energy, we also calculated the cross-correlation between them as 
\begin{equation}
 C_{E_0,E_1}=\langle E_0E_1 \rangle -\langle E_0\rangle\langle E_1\rangle.
\end{equation}
  As expected the results shown in Fig.\ \ref{fluctuation} indicate that $E_0$ and $E_1$ are anti-correlated. It also provides a signature of the coil-globule transition in both models, especially for the higher $\kappa$ values. However, $C_{E_0,E_1}$ also fails to capture any signature of the transition to the knotted phase. Thus, we call for an analysis deployed specifically to knots in the polymer. }

\begin{figure*}[t!]
\includegraphics[width=0.9 \textwidth]{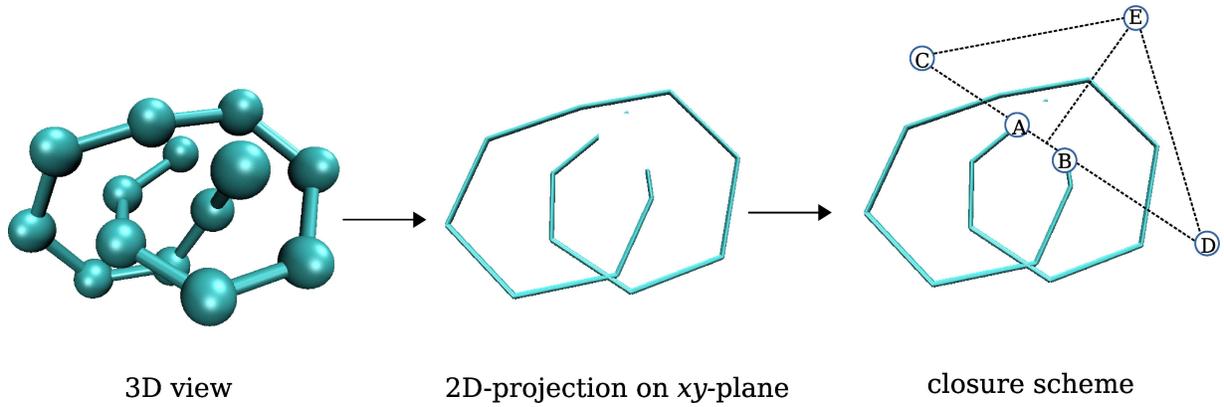}
\caption{{Illustration of the closure scheme to identify the knot type in a polymer. Left panel represents the 3D conformation of a knotted ($3_1$) polymer. Central panel shows the 2D-projection of the same polymer on $xy$-plane. The right panel demonstrates the closure applied on the 2D-projection to make the open polymer a closed one. } }
\label{closure-scheme}
\end{figure*}

\begin{figure}[t!]
\includegraphics[width=0.475 \textwidth]{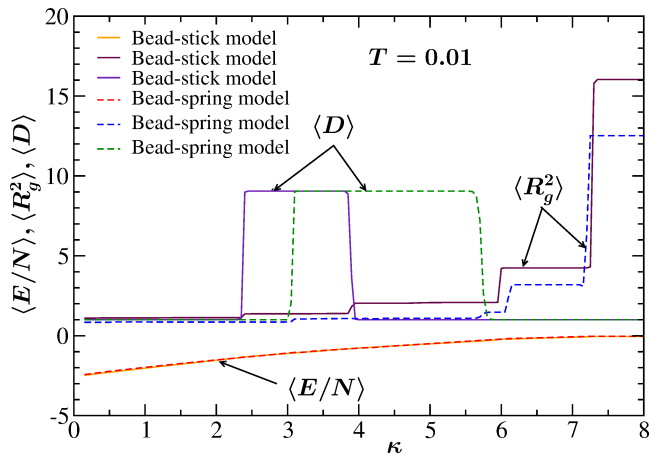}
\caption{Variation of the energy density $\langle E/N\rangle$, the squared radius of gyration $\langle R_g^2 \rangle$, and the knot parameter $\langle D \rangle$ as a function of the 
bending stiffness $\kappa$ for a fixed temperature $T=0.01$ for both models with the choice of the ratio $r_b/r_{\rm{min}}=0.891$ and $N=14$. }
\label{parameter_vs_kappa}
\end{figure}
\par
{In a mathematical sense knots are only defined for closed curves as for the schematics shown in Table\ \ref{tab1_5}. An open
polymer can satisfy the mathematical definition of a knot only when the termini are closed virtually. For that we follow Ref.\ \cite{virnau2010} and first project the polymer conformation on 
a 2D plane as illustrated in Fig.\ \ref{closure-scheme} for a conformation with a $3_1$ knot. One can notice that the mere 2D-projection (say on the $xy$-plane) yields only one crossing. A direct closure of the termini A and B would also not yield any additional crossing. Therefore, one needs a special closure scheme as demonstrated in the right most panel of Fig.\ \ref{closure-scheme}. There we connect the termini A and B by a straight line, which is then extended in both directions to get two new virtual points C and D located far away from all the monomers. Following that we create another virtual point E, far away from all the monomers, on the perpendicular bisector of the line AB. The polymer is now closed via straight lines connecting E to C and D, respectively. The resulting closed curve now has two additional crossings making the total number of crossings to be three. The closure is
only applied during the measurement of the knot type and does not influence the simulation itself. The details of this closure prescription can be found 
in Refs.\ \cite{virnau2005knots,virnau2010,marenz2016knots,janke2016stable}.}
\par
{A knot type is denoted as $C_n$ where the integer $C$ counts the minimum number of crossings and the subscript $n$ distinguishes topologically different knots with the same number of crossings \cite{kauffman1991}.
In our study, once the closure is applied the knot type of the polygonal line describing the polymer is determined in the following way. First we identify the crossings and then determine the corresponding Alexander polynomial \cite{kauffman1991}. In order to avoid unwanted prefactors of the Alexander polynomial $\Delta(t)$, we calculate a variant of it given as
\begin{equation}
 \Delta_p(t)=\lvert \Delta(t) \times \Delta(1/t) \rvert,
\end{equation}
evaluated at $t= -1.1$. Thus we define the knot parameter as $D\equiv \Delta_p(-1.1)$. $D$ is also a knot invariant which implies that different polygonal lines with the same knot type correspond to the same $D$.  However, it is not unique as the underlying Alexander polynomial is not unique [e.g.,
$D(5_1)=D(10_{132})$]. Nevertheless, it is sufficient to
distinguish between the simple knots observed in this work. Once the knot parameter 
$D$ is found for a polymer conformation one can assign the knot type $C_n$ from a list of possible values of $D$ for simple knots, as presented in Table\ \ref{tab1_5}.}

\par
The estimated average of the knot parameter $\langle D \rangle$ for the two models is shown in Figs.\ \ref{parameters}(a) and (b), respectively. For the bead-stick and the bead-spring polymer with $\kappa=3.0$ and $4.0$, respectively, one can clearly see that at low temperature $\langle D \rangle$ coincides with the value of $D=9.05462$ that specifies a trefoil knot ($3_1$) and at higher $T$, it drops 
down to $1$ that corresponds to an unknotted polymer. Thus, undoubtedly the knot parameter is the distinguishing parameter we should be exploring in this work. This can also be appreciated from 
the plots in Fig.\ \ref{parameter_vs_kappa} showing comparative variation of $\langle E/N\rangle$, $\langle R_g^2 \rangle$, and $\langle D \rangle$ as a function of $\kappa$ for the temperature fixed to our lowest value of $T=0.01$. There also indeed $\langle D \rangle$ provides the most convincing picture for the transition to the knotted phase K$3_1$ for both models. 
\begin{figure}[t!]
\includegraphics[width=0.475 \textwidth]{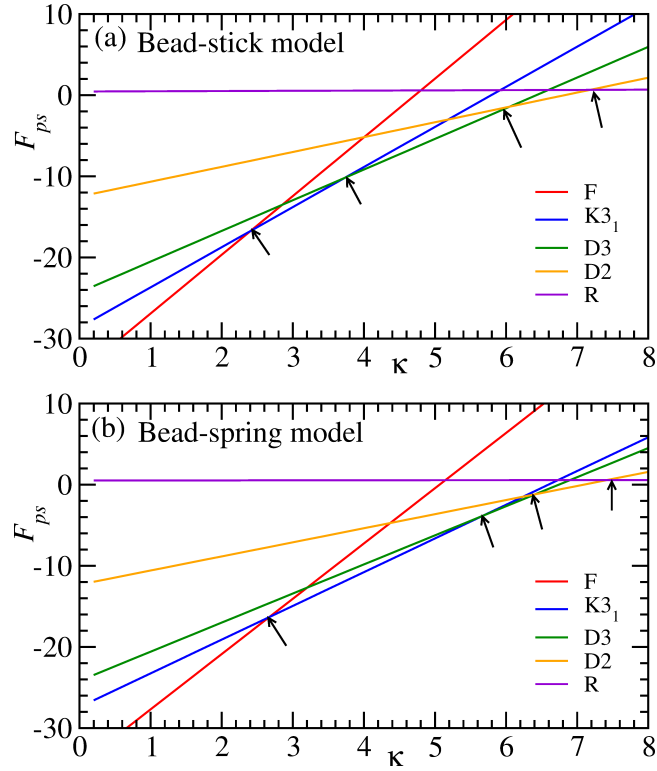}
\caption{Variation of the calculated pseudo free energy $F_{ps}$ as a function of the bending stiffness $\kappa$ at a fixed temperature $T=0.01$. Results for typical conformations identified from the phase diagrams presented in Fig.\ \ref{Full-PD} for (a) bead-stick model and (b) bead-spring model are presented. The arrows there mark the position where the conformation with the minimum energy switches from one structure to the other.}
  \label{Pseudo-fenergy}
\end{figure}
\par
Observation of knotted structure is not really new, however, in the past the knotted structures 
found were by chance and hence were mostly observed in the coiled and globular states. Here, the 
full phase diagrams in Fig.\ \ref{Full-PD} indicate that the knotted structures are the stable phases for intermediate values of bending stiffness, especially at low temperatures. This fact raises the question whether there are any entropic contributions to these stable knotted phases. We investigate this in the following empirical approach. We pick up typical 
conformations (F, K$3_1$, D3, D2, and R) at the lowest temperature $T=0.01$ which can be identified from  Fig.\ \ref{Full-PD} for both the bead-stick and the bead-spring model. Now keeping their 
morphology intact we calculate the total energy $F_{ps}$ of each of them just by varying the bending stiffness $\kappa$ using the Hamiltonian in Eq.\ \eqref{hamiltonian}. Since this is done at $T=0.01$ and assuming that the entropic contributions are negligible, it can be considered that one calculates virtually the free energies 
of the respective conformation while changing $\kappa$. Hence, $F_{ps}$ could be termed as the pseudo free energy of those conformations. 

\par
In Figs.\ \ref{Pseudo-fenergy}(a) and (b) we present the variation of $F_{ps}$ with $\kappa$ at $T=0.01$ for a set of typical conformations, respectively for the bead-stick and bead-spring model. From the plot one can easily identify which conformation has the minimal $F_{ps}$ at a particular value of the stiffness $\kappa$. For example, when $\kappa=2.0$ for both models the frozen conformation (F) has the lowest $F_{ps}$. Similarly, for $\kappa > 7$ the rod-like (R) conformation has the minimal $F_{ps}$. This observation is in concurrence with the full phase diagrams presented in Fig.\ \ref{Full-PD} for both models. If one starts at $\kappa=0$ and moves on with increasing $\kappa$, at some value of $\kappa$ the K$3_1$ knot takes over the frozen conformation as the conformation with minimum $F_{ps}$. This crossover or switching (marked by the arrows in the plots) to different 
conformations having the minimum $F_{ps}$ happens four times along the $\kappa$ axis for both models. Interestingly, these crossover points along $\kappa$ match quite well with the phase boundaries one observes in the full phase diagrams in  Fig.\ \ref{Full-PD}. This confirms that for all these conformations at low temperature the entropic contribution is indeed negligible. 
\begin{figure}[b!]
\includegraphics[width=0.48 \textwidth]{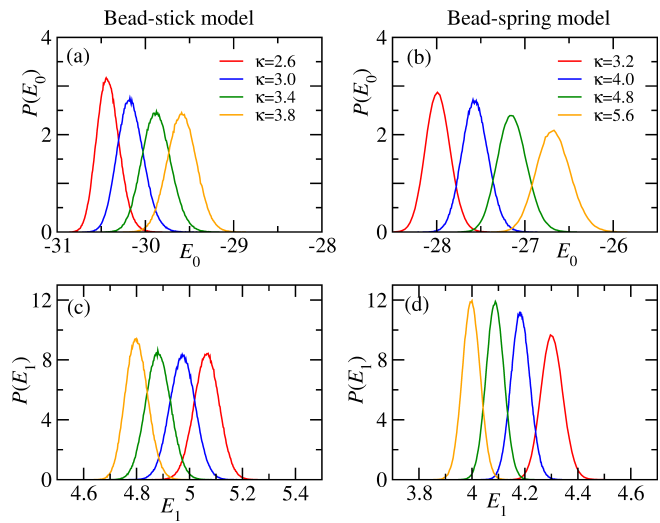}
\caption{Probability density of the base energy $E_0$ in the knotted phase for different values of the bending stiffness $\kappa$ for (a) bead-stick model and (b) bead-spring model at a temperature $T=0.01$. Plots in (c) and (d) show the corresponding probability densities of the energy $E_1$ ($=E_{\rm{bend}}/\kappa$) for the two models. All the results are for the ratio $r_b/r_{\rm{min}}=0.891$ and $N=14$.}
  \label{Knot-energy-dist}
\end{figure}
\par
The variations of the mean energy in Fig.\ \ref{parameter_vs_kappa} for both models are consistently overlapping with each other and are almost indistinguishable. Thus the wider range for stable knotted conformations in the bead-spring model 
should be attributed to the interplay of the base energy $E_0$ and the bending energy $\kappa E_1$. 
To have an idea about this interplay, we show in Figs.\ \ref{Knot-energy-dist}(a) and (c) for the bead-stick model, the probability density of $E_0$ and $E_1$, respectively, for four different values of $\kappa$ within the knotted phase at $T=0.01$. The corresponding plots for the bead-spring model are presented in Figs.\ \ref{Knot-energy-dist}(b) and (d). 
$E_0$ for the bead-stick model is the nonbonded energy $E_{\rm{nb}}$ described in 
Eq.\ \eqref{potential_OLM}. For the bead-spring model, $E_0$ also consists of the bond energy $E_{\rm{FENE}}$ [as in Eq.\ \eqref{FENE}], in addition to $E_{\rm{nb}}$. $E_1$ in both models correspond to $E_{\rm{bend}}/\kappa$. Thus $E_1$ accounts for the relative orientation of the bonds along the length of the polymer, i.e., the factor $\sum_{i=1}^{N-2}(1-\cos \theta_i)$ in Eq.\ \eqref{stiff}. From Figs.\ \ref{Knot-energy-dist} (a) and (b) it can be observed that the peak of the distribution of $E_0$ shifts to the right with increase of $\kappa$ for both models. On the other hand, from Figs.\ \ref{Knot-energy-dist} (c) and (d) it is observed that this trend is opposite for $E_1$, albeit the $E_{\rm{bend}}$ anyway increases as $\kappa$ increases. Thus for both models it is apparent that a decrease in $E_1$ is paid off by the increase in $E_0$. For the bead-spring model the increase in $E_0$ per unit change in $\kappa$ is $\approx 0.5$, which is smaller than the corresponding variation $\approx 0.83$ for the bead-stick model. Similarly, the corresponding decrease in $E_1$ per unit change in $\kappa$ within the knotted phase is 
smaller in the bead-spring model ($\approx 0.125$) than in the bead-stick model ($\approx 0.25$). This 
difference comes from the fact that in the bead-spring model since the bond length is not fixed a variation in the bond lengths may also give rise to an overall better orientation of the bonds such that $E_1$ is decreased. At very large $\kappa$ as the overall $E_{\rm{bend}}$ becomes large and thereby mild bond orientation is not enough to stabilize the structures, and hence bent structures appear and the knotted phase vanishes. For the bead-stick model the allowed range of $\kappa$ is potentially small since the change in bond orientation, i.e., decrease in $E_1$ is only possible due to a pure bond rotation. This provides an intuitive argument why the knotted phase is much wider in the bead-spring model than in the bead-stick model.

\subsection{Existence of knots while varying $r_b/r_{\rm{min}}$}
\begin{figure}[b!]
\includegraphics[width=0.45 \textwidth]{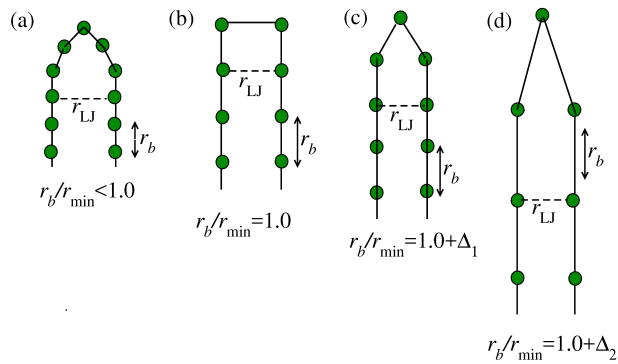}
\caption{Schematic diagram showing the possible perfectly bent structures one can observe in a semiflexible polymer model with $4$ different choices of the ratio $r_b/r_{\rm{min}}$ as mentioned. Here $0<\Delta_1< \Delta_2$.}
\label{schematic}
\end{figure}

From the results obtained in the previous subsection with the ratio $r_b/r_{\rm{min}}=0.891$ in both models we conclude that the formation of a stable 
knotted phase at low temperatures is guided by the interplay of the base energy $E_0$ (where $E_{\rm LJ}$ is the sole respectively major contribution for the bead-stick or bead-spring model) and the bending energy $E_{\rm{bend}}$. 
Thus for $r_b/r_{\rm{min}}=0.891$ one obtains frozen conformations 
F where the energy minimization due to the nonbonded contacts can easily overcome the required bending energy penalty (for F there are a number of bends along the chain that have a bending angle $\theta \approx \pi/2$) for such conformations. As the stiffness $\kappa$ increases, naturally the number of bends along the chain shall decrease which may give rise to bent conformations like D3 and D2. However, for this specific choice of $r_b/r_{\rm{min}}=0.891$  it is observed that for intermediate values of the stiffness (for both models) knotted conformations are observed. A knotted conformation, like $3_1$ has much less number of severe bends (bending angle $\theta \ll \pi/2$) than a frozen conformation, but has sufficient nonbonded contacts courtesy to the crossing or overpassing of the chain onto itself to fulfill the topology of a knot. At even larger values of $\kappa$ the nonbonded LJ interaction due to the knot topology is not enough to overcome the bending penalty, and thus bent structures like D3 or D2 becomes the stable ones. In such a conformation an energy gain is achieved via the nonbonded LJ contacts of the opposite strands. Now, it is easy to perceive that this strength of the LJ contact is maximum when the distance $r_{\rm{LJ}}$ between the strands coincides with $r_{\rm{min}}$ of the model (see Fig.\ \ref{schematic} for the definition of $r_{\rm{LJ}}$ in this context). On the other hand, the possible value of this $r_{\rm{LJ}}$ is correlated with the equilibrium bond length $r_b$ of the conformation (again see Fig.\ \ref{schematic} to correlate $r_b$ with $r_{\rm{LJ}}$). This leads to the inference that the stability of the 
bent conformations is dependent on the ratio $r_b/r_{\rm{min}}$. Since, the existence of the knotted phases is dependent on its 
energetic competition with the bent phases, thus in turn the very existence of the knotted phases is practically dependent on this ratio $r_b/r_{\rm{min}}$. 

\par
\begin{figure}[t!]
\includegraphics[width=0.425 \textwidth]{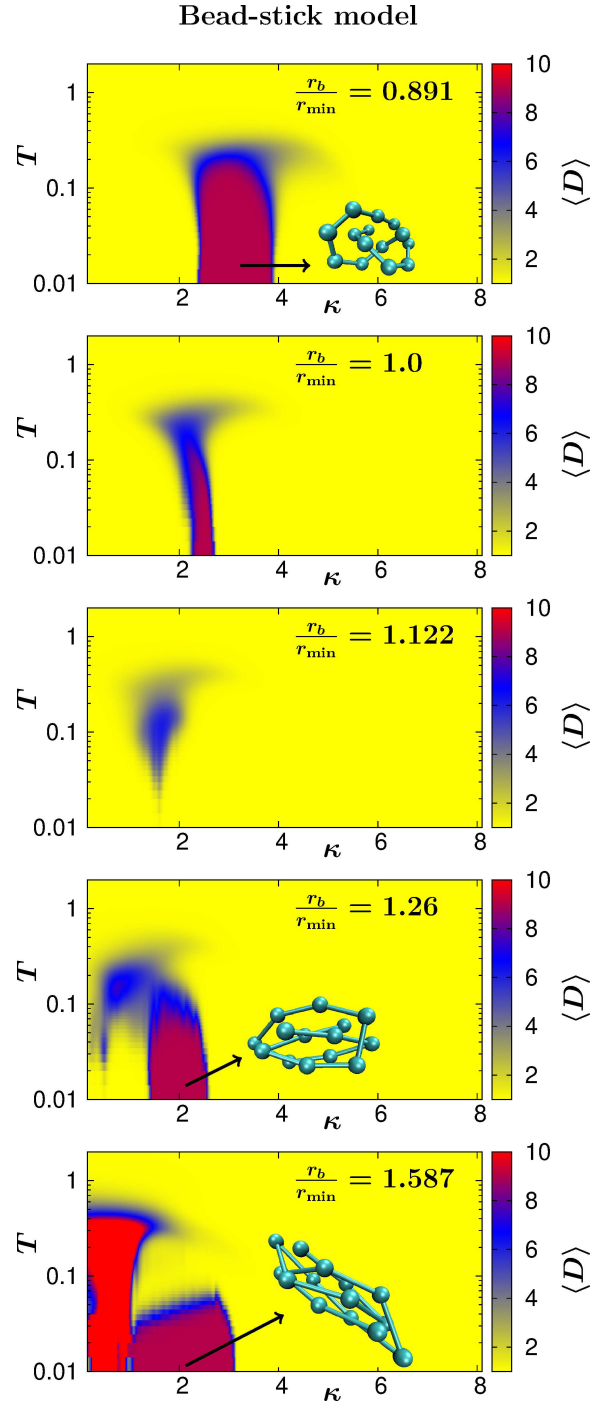}
\caption{Phase diagram in ($T,\kappa$) plane with the knot parameter $\langle D \rangle$ as the order parameter for a semiflexible polymer with different choices of the ratio $r_b/r_{\rm{min}}$ using a bead-stick model. {The snapshots represent typical polymer conformations in the stable knotted phase having trefoil knots ($3_1$) for the respective choices of $r_b/r_{\rm min}$.} All the results are for a chain length $N=14$. }
 \label{Knot-PD-diff-ratio-Bst}
\end{figure}
\begin{figure}[t!]
\includegraphics[width=0.425 \textwidth]{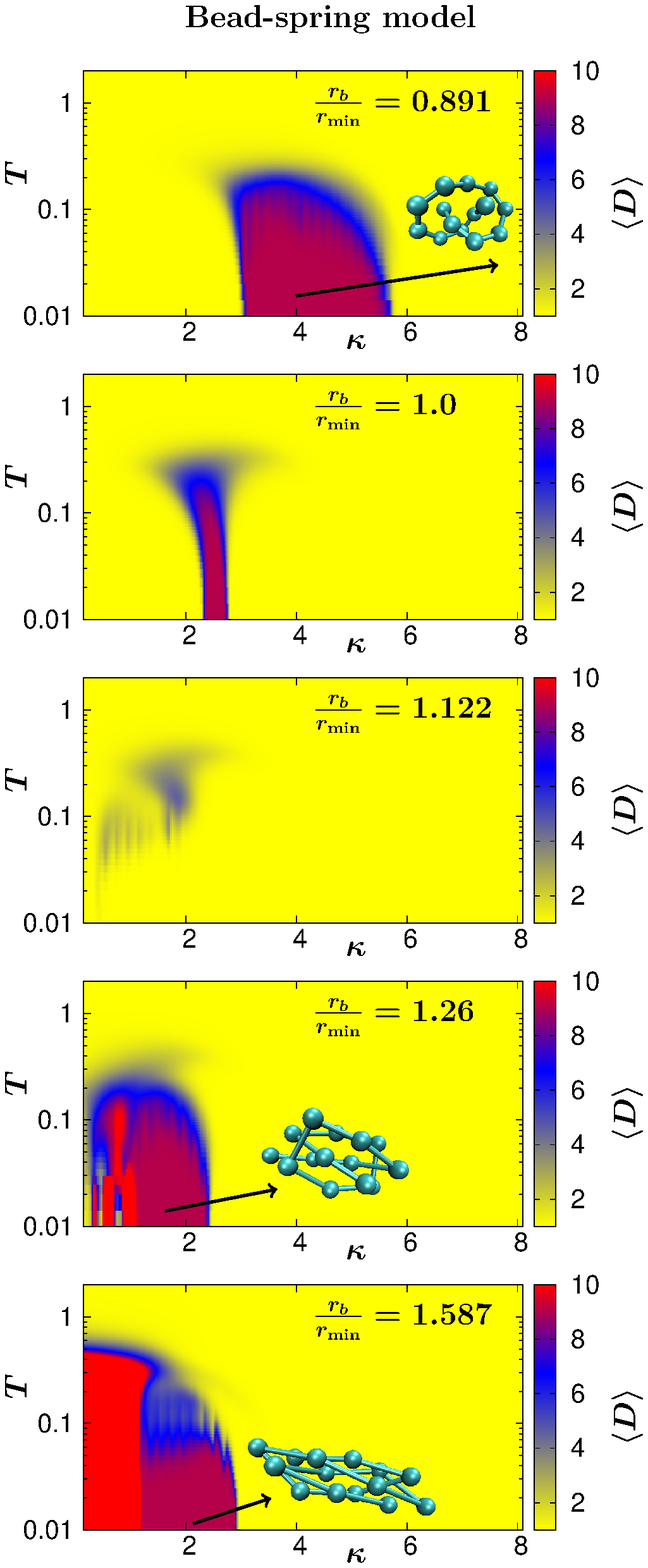}
\caption{ Same as Fig.\ \ref{Knot-PD-diff-ratio-Bst} but for the bead-spring model with $N=14$.}
 \label{Knot-PD-diff-ratio-Bsp}
\end{figure}

Before we move on to explore the existence of knots in both models for various choices of the ratio $r_b/r_{\rm{min}}$, in Figs.\ \ref{schematic}(a)-(d) we illustrate 
our speculation about the stability of a bent conformation with D2 as an example. The schematic diagram shows the possible two-dimensional projection of a stable D2 conformation for four 
different typical choices of $r_b/r_{\rm{min}}$. In (a) we have drawn such a schematic for $r_b/r_{\rm{min}} < 1$. For a short chain of length $N=14$, in  this case, since $r_b < r_{\rm{min}}$, to have the maximum nonbonded LJ interaction, i.e., to have $r_{\rm LJ}=r_{\rm{min}}$ the bending tip must include several monomers which in turn leave only few monomers to have a real gain in energy due to LJ contact with the opposite strands. Thus, for intermediate values of the bending stiffness the bent conformations are unstable compared to a trefoil knotted ($3_1$) conformation (see the typical conformations in Figs.\ \ref{Knot-PD-diff-ratio-Bst} and \ref{Knot-PD-diff-ratio-Bsp} for $r_b/r_{\rm{min}} =0.891$). For the case of  $r_b/r_{\rm{min}}= 1$ and $r_b/r_{\rm{min}}= 1+\Delta_1$ drawn respectively in Figs.\ \ref{schematic}(b) and (c), the minimum number of monomers 
involved in the bending to form two strands are respectively two and three (in general, they must be fewer than in the case of $r_b/r_{\rm{min}} < 1$). Thus, more monomers can stay on the strands which can now lie easily at a distance $r_{\rm LJ}=r_{\rm min}$ thus minimizing the energies at even intermediate values of the bending stiffness $\kappa$. Hence, it seems that for such cases the bent conformations are always favorable over the simple knotted structure $3_1$, possible for relatively short chain length $N$. However, this is restricted by the value $\Delta_1$. Now let us compare the cases in (b) and (c). In (b) the full turning of the polymer involves two bendings (with $\theta_i=\pi/2$) which accounts for a bending energy $2\kappa$. In this case for a polymer of length $N$ the total number of nonbonded contacts will be $(N-2)/2$ which 
accounts for an energy gain of $-(N-2)\epsilon/2$. For the case in (c) the gain in energy due to nonbonded contacts would be the same as in (b), i.e., $-(N-2)\epsilon/2$. However, in this case there are three bends for the full turning of the polymer. Thus in this case the relative orientations of these three bonds involved in the turning would decide the total bending energy penalty. Now, if $\sum_i (1-\cos \theta_i) < 2$ then the conformation in (c) will be even more stable than the corresponding structure in (b). This is dependent on the value of $\Delta_1$. For smaller values of $\Delta_1$, the condition $\sum_i (1-\cos \theta_i) < 2$ is satisfied, and thus the bent structures are even stabler and one would not expect to observe a simple knotted phase in the phase diagram. However, if $\Delta_1$ is very large then $\sum_i(1-\cos \theta_i) > 2$ and the bent structure in (c) gradually becomes less stable compared to (b) and eventually compared to even a trefoil ($3_1$)  knotted structure {(see the typical conformations in Figs.\ \ref{Knot-PD-diff-ratio-Bst} and \ref{Knot-PD-diff-ratio-Bsp} for $r_b/r_{\rm{min}} =1.26$ and $1.587$)}. Schematic for such a case, i.e., with $r_b/r_{\rm{min}}= 1+\Delta_2$ (where $\Delta_2 > \Delta_1, ~r_b \gg r_{\rm {min}}$) is shown in (d). There one can easily notice that the apex angle $\theta$ approaches $\pi$, thus making the overall bending energy larger again. In such a situation thus we speculate that at lower or intermediate values of $\kappa$ simple knotted structure like $3_1$ would be again favorable. From the above heuristic arguments, we conjecture that for polymers of short length except for a small window of the ratio $r_b/r_{\rm{min}} \in [1,1+\Delta_1]$ one would expect to observe a knotted phase at lower or intermediate values of the bending stiffness. 

\par
To check the validity of the above arguments and how the existence of the knotted phase gets affected by the ratio $r_b/r_{\rm{min}}$, we perform simulations with both the bead-stick and bead-spring model for four other choices of  $r_b/r_{\rm{min}}=1.0,~2^{1/6}~(\approx 1.122),~2^{2/6}~(\approx 1.26),~ \rm{and}~2^{4/6}~(\approx 1.587)$. Note that this ratio has a lower bound decided by the 
fact that $r_b$ cannot be less than the diameter $\sigma$ of the monomer beads. This puts the lower limit to the ratio $r_b/r_{\rm{min}} = 2^{-1/6}\approx 0.891$ below which we do not perform any simulations. There is no strict upper bound on $r_b/r_{\rm{min}}$ but we go up to the value $1.587$ beyond that both models show no condensed structure at all for a polymer of length $N=14$. 
\par
Figure\ \ref{Knot-PD-diff-ratio-Bst} shows the results for the bead-stick model which illustrate how the existence of a stable knotted phase gets influenced by the ratio $r_b/r_{\rm{min}}$. The phase diagram in the ($T,\kappa$) plane is constructed as a surface plot using the knot parameter $\langle D \rangle$. There one clearly sees a knotted phase at low temperatures and intermediate bending stiffness $\kappa$ for all ratios except $r_b/r_{\rm{min}}=1.122$. For the 
case $r_b/r_{\rm{min}}=1.0$ the region of knotted phase is very narrow. This is in concurrence 
with our speculation that the bent structures are favorable over the knotted ones for the cases 
presented in Figs.\ \ref{schematic}(b) and (c). We have checked that for  $r_b/r_{\rm{min}}=1$ and $1.122$ alternative structures which appear are D3 and D2. The slightly higher values of  $\langle D \rangle$ marked by the blue spot in the phase diagram for $r_b/r_{\rm{min}}=1.122$ is 
 due to the presence of few knotted structures mixed with the simple globule. These knots are 
 not stable knots but are formed by chance and are hence of the kind of knotted structures which were reported in the past. Note that since the chain length is relatively short it is impossible to observe a wide variety of knotted structures. In fact in all the cases the observed knots  correspond mostly to the trefoil knot $3_1$ characterized by the $D=9.05462$ (see Table\ \ref{tab1_5}). This can be identified by the red colored region in Fig.\ \ref{Knot-PD-diff-ratio-Bst}. For $r_b/r_{\rm{min}}=1.587$ one also notices an orange region at very low $\kappa$ up to relatively high $T$. We caution the reader that this does not correspond to the $8_{19}$ knot with $D=9.72667$, since a knot with $8$ crossings is impossible for a chain length of $N=14$. Rather this region corresponds to a mixed phase of $4_1$ knots (having $D=25.09099$) and unknotted conformations. These knots are qualitatively different. They originate in the frozen amorphous (or glass-like) state of the polymer and are highly unstable. Thus a small perturbation is sufficient to unknot them. The other knot $3_1$ is a toroidal knot which reduces the bending energy and is thus stable.
 \begin{figure}[t!]
\includegraphics[width=0.45 \textwidth]{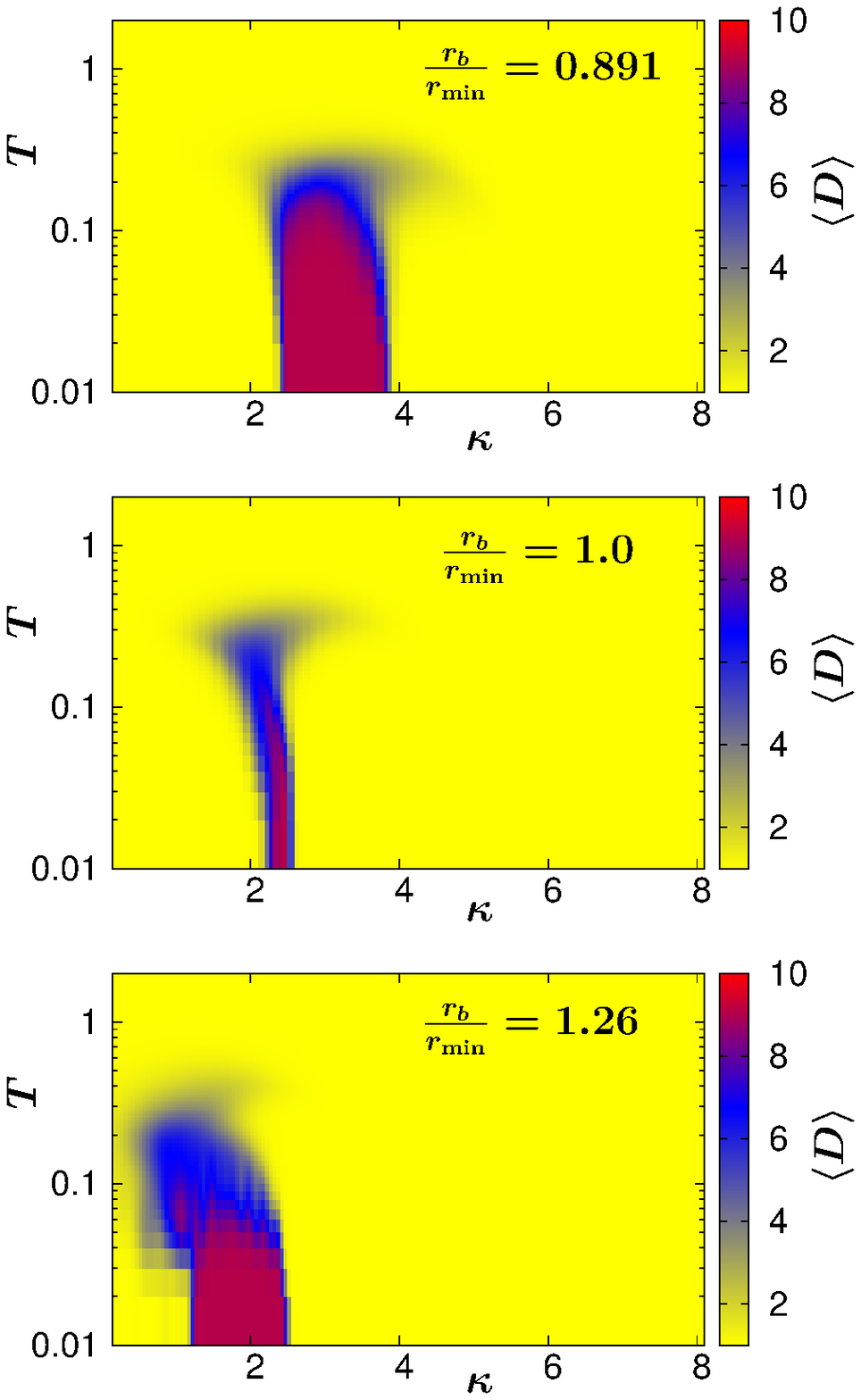}
\caption{Phase diagram in ($T,\kappa$) plane with the knot parameter $\langle D \rangle$ as the order parameter for a semiflexible polymer with different choices of the ratio $r_b/r_{\rm{min}}$ using a bead-spring model with the spring constant $K=297.5$. The results are for a chain length $N=14$.}
\label{Knot-PD_K297}
\end{figure}
 \par
 Similar observation can be made from the results of the bead-spring model presented in  Fig.\ \ref{Knot-PD-diff-ratio-Bsp} for the same choices of the ratio $r_b/r_{\rm{min}}$. {In addition one can notice that for all the ratios the width of the knotted phase in the bead-spring model is larger compared to the corresponding width in the bead-stick model. Here, also the orange region for $r_b/r_{\rm{min}}=1.587$ at low $\kappa$ and covering even relatively higher $T$ corresponds to the mixed phase comprising unknotted and $4_1$ knotted structures. Another observation which is in place for both the models is that in contrast to the generic ``spherical'' knot that one obeserves for $r_b/r_{\rm min}=0.891$, for 
 the highest value of $r_b/r_{\rm{min}}=1.587$ the knotted conformation looks similar to a bent structure. A careful look in comparison with the bent conformations observed with  $r_b/r_{\rm{min}}=0.891$ in Fig.\ \ref{Full-PD} would reveal that the internal structures are different. In case of $r_b/r_{\rm{min}}=1.587$ the strands penetrate each other giving rise to a ``flat'' knotted structure thereby costing bending energy but gaining energy due to additional LJ contacts. In contrast, for $r_b/r_{\rm{min}}=0.891$, the strands in the bent structures are almost parallel to each other.}
 \par
The observed realization of knotted phases in the bead-spring model raises the question why they were not noticed by Seaton \textit{et al.}\ \cite{seaton2013flexible} with their bead-spring model. 
The details of this model are discussed in the Appendix. The main difference between their bead-spring model and the model we used is the consideration of the bond energy $E_{b}$. 
In their case the bonded monomers in addition to a FENE potential also interact via a LJ kind of potential. This makes the effective spring constant that takes care of the 
elasticity of the bonds much larger, as is shown via the harmonic approximation in the Appendix. {Instead of simulating exactly the model of Seaton \textit{et al.} \cite{seaton2013flexible} we choose to simulate our 
bead-spring model with a spring constant $K=297.5$ in Eq.\ \eqref{FENE} to be equal to the effective spring constant of the bonds $K_{\rm eff}$ given in Eq.\ \eqref{K_eff}.} 

\par
The results for the bead-spring model with $K=297.5$ in the FENE bonds for different choices of the ratio $r_b/r_{\rm{min}}=0.891$, $1.0, \rm{and}~1.26$ are presented in 
Fig.\ \ref{Knot-PD_K297}. In this case also, one can clearly see that 
for  $r_b/r_{\rm{min}}=1.0$ the knotted phase region is very narrow on the ($T,\kappa$) plane and that it is significantly wider for the cases when $r_b/r_{\rm{min}} =0.891$ and $1.26$. This again is in accordance with our speculations. Thus they are qualitatively similar to the results presented in Fig.\ \ref{Knot-PD-diff-ratio-Bsp} where $K=40$. However, closer inspection reveals that the ranges of $\kappa$ over which one sees the knotted phase are $[2.4,3.8]$ and $[1.2,2.4]$, respectively, 
for $r_b/r_{\rm{min}}=0.891$ and $1.26$,  which are smaller than the corresponding ranges for the bead-spring model with $K=40$. On the contrary, these ranges almost coincide with those we found for the respective values of $r_b/r_{\rm{min}}$ using the bead-stick model presented in Fig.\ \ref{Knot-PD-diff-ratio-Bst}. Such a good match with the bead-stick model shows that using a high value $K=297.5$ makes the FENE bonds in the bead-spring model almost as rigid as in the bead-stick case. The realization of a knotted phase in Fig.\ \ref{Knot-PD_K297} points to the fact that one would have also observed a knotted phase in the model used by 
Seaton \textit{et al.} had they used the ratio $r_b/r_{\rm{min}}=0.891$ and $1.26$. In their study \cite{seaton2013flexible} they used $r_b/r_{\rm{min}}=1.0$ for which anyway we 
expect the knotted region to be very narrow. Also, the lowest temperature down to which they simulated was $T=0.03$, for which the chance of detecting the stable knotted phase is really poor. To substantiate our finding, as a step further we simulated even a longer chain ($N=28$) using our bead-spring model 
with $K=297.5$ and $r_b/r_{\rm{min}}=1.0$ (see the Appendix) which in principle is equivalent to the bead-spring model of Ref.\ \cite{seaton2013flexible}. There also we do not find any stable knotted phase.

 \subsection{Richer knotted phase behavior for longer polymers}
So far all the results we have presented are for a polymer of length $N=14$. There we essentially found the presence of a specific knot type $3_1$, the trefoil knot with $D=9.05462$. This observation of a single knot type is due to the short length which does not allow too many crossings. As expected if the length increases the possibility of having many crossings increases which should give rise to a rich variety of knotted structures. It is also quite intuitive that as the length of the polymer increases the chances of forming knots will be higher. This can be compared with the ease with which one can tie a knot if the given thread is longer. This 
 could explain the formation of the knots which are formed by chance in the globular or coiled phase. Nevertheless, we expect that the likelihood of finding low-temperature stable knotted phases will also increase. 
 \begin{figure*}[t!]
\includegraphics[width=0.425 \textwidth]{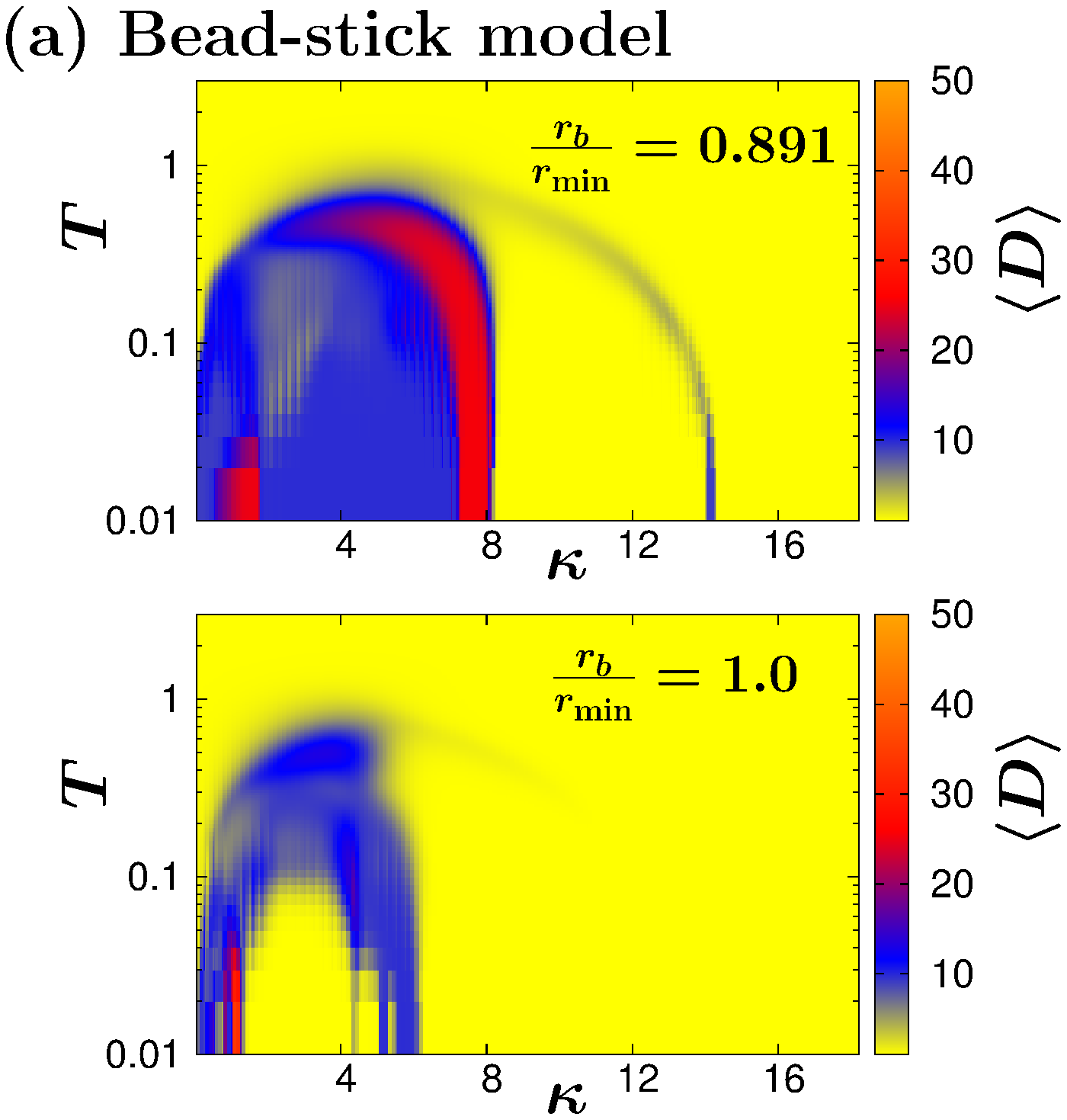}
~~~\includegraphics[width=0.425 \textwidth]{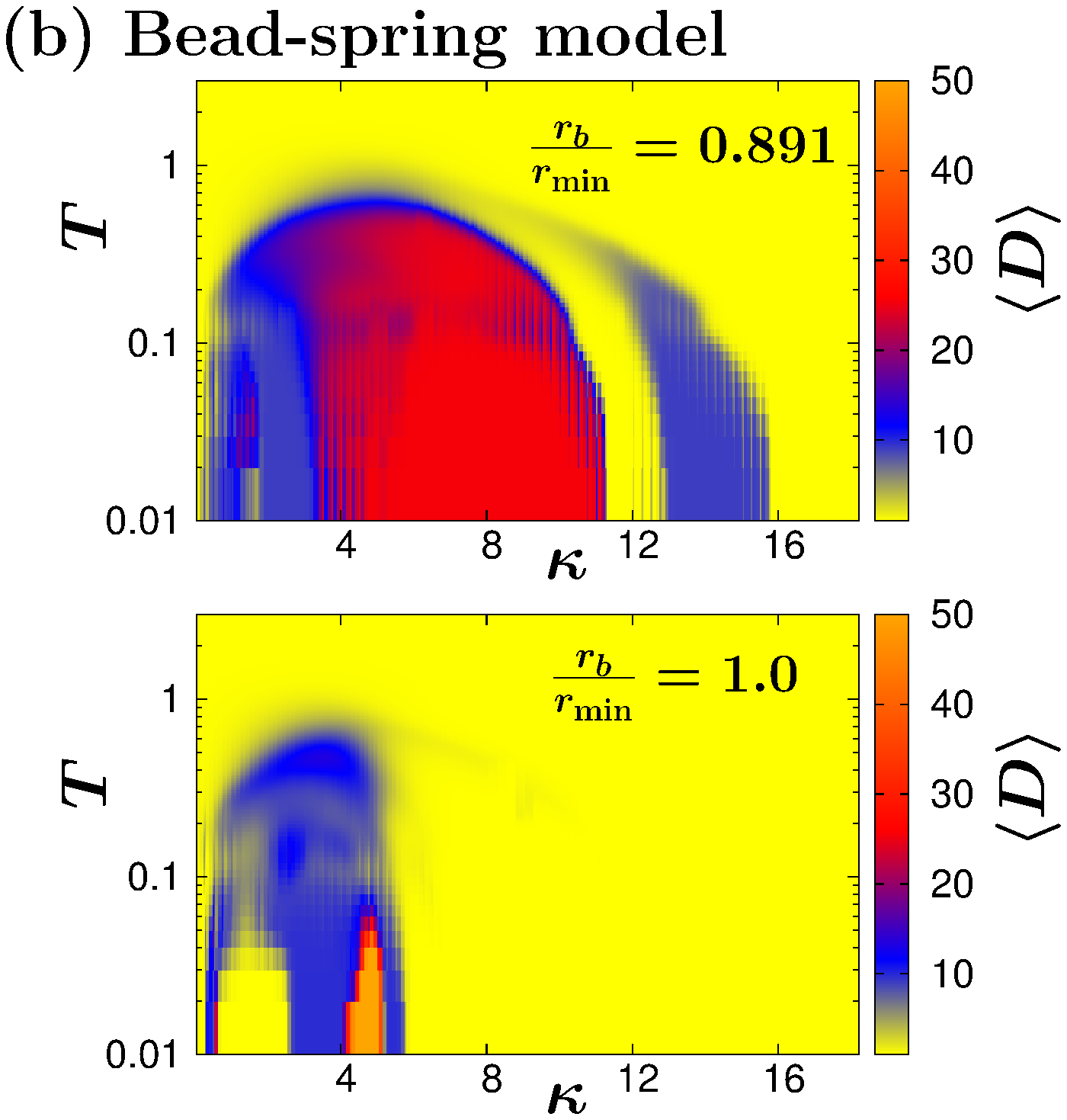}\\
\caption{Phase diagram in ($T,\kappa$) plane with the knot parameter $\langle D \rangle$ as the order parameter for a semiflexible polymer of length $N=28$ using (a) the bead-stick model and (b) the bead-spring model. For both models results for two choices of the ratio  $r_b/r_{\rm{min}}=0.891$ and $1.0$ are presented.}
\label{Knot-PD-N28}
\end{figure*}
\par
To investigate the existence of knots in longer chains we now simulate a polymer of length $N=28$ for both models. This choice of $N$ can be compared with some recent experimental and numerical studies of synthetic polymers adsorbed on a surface in vacuum \cite{forster2014structure,forster2014polymer}. For each case we pick two different values of $r_b/r_{\rm{min}}=0.891$ and $1.0$, for which we speculated to have respectively presence and absence (or a very narrow range) of stable knotted phases. The corresponding phase diagram for the existence of knots are 
shown in Figs.\ \ref{Knot-PD-N28}(a) and (b), respectively, for the bead-stick and the bead-spring model. As expected for the ratio $r_b/r_{\rm{min}}=0.891$ both models exhibit a stable knotted phase over a wide range of low to intermediate bending stiffnesses $\kappa$. For the case with $r_b/r_{\rm{min}}=1.0$, however, both models show a much smaller window for the knotted phase as observed for the $N=14$ case. 

\par
\begin{figure*}[t!]
\includegraphics[width=0.45 \textwidth]{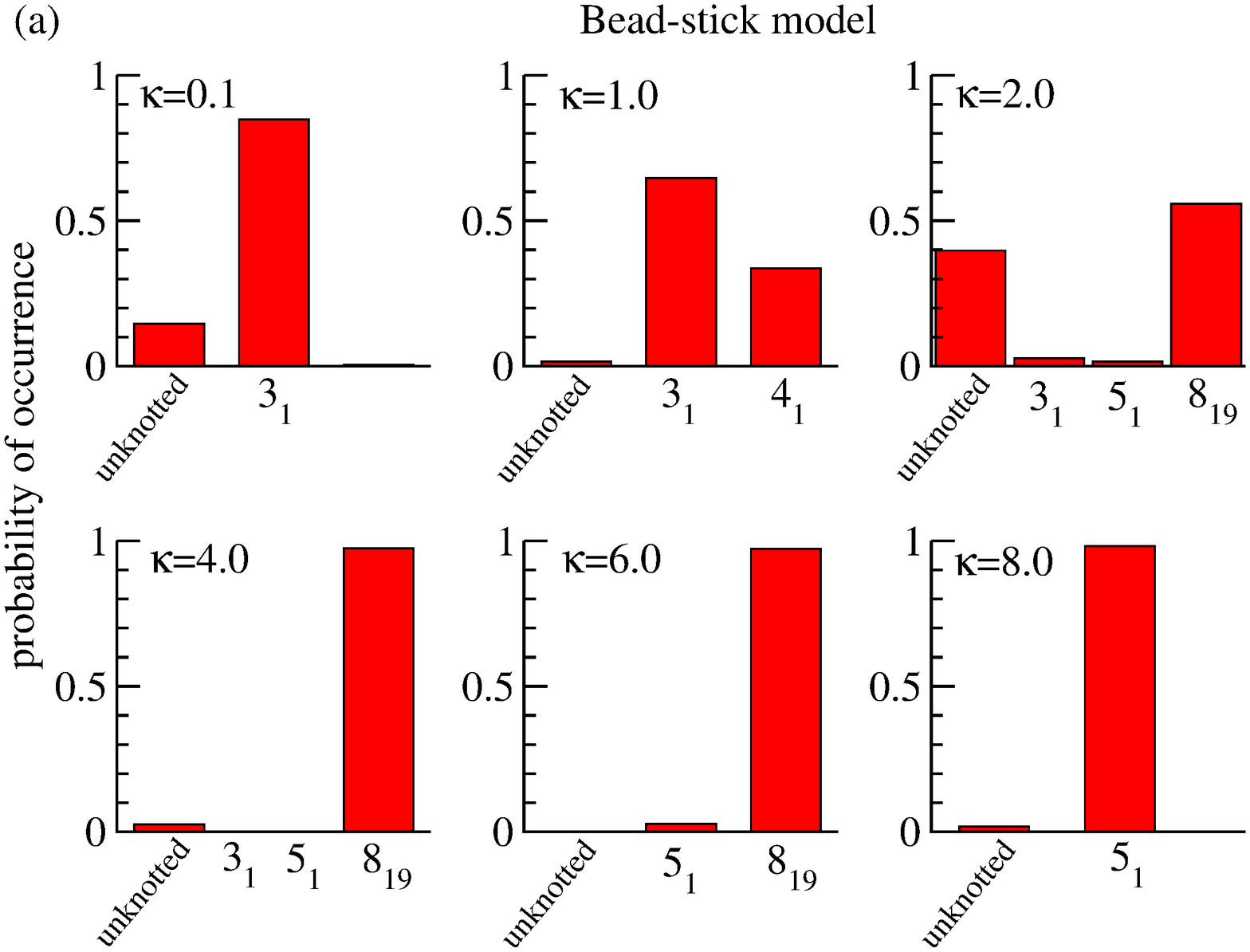}~~~~
\includegraphics[width=0.45 \textwidth]{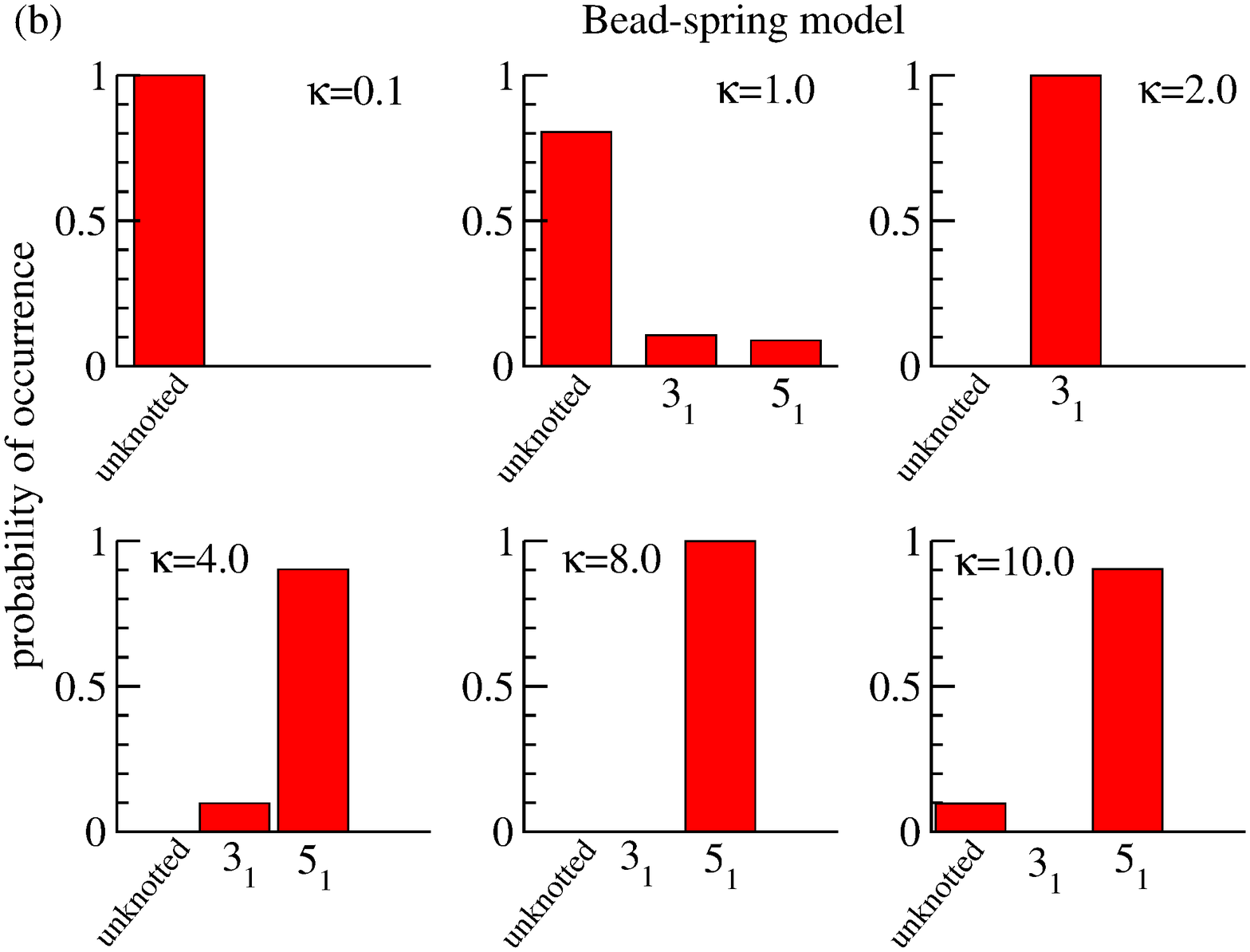}
\caption{Probabilities of occurrence of different knot types at $T=0.01$ and at different bending stiffness $\kappa$ for the two models. The results are for a polymer of length $N=28$ with the ratio $r_b/r_{\rm{min}}=0.891$.}
\label{Knot-Prob_N28}
\end{figure*}

From the range over which the estimated knot parameter $\langle D \rangle$ varies for both models it is clear that there exist different knot types indicating a much richer knotted phase behavior compared to the $N=14$ case. However, it is not possible to have an idea about the different types from these phase diagrams. {Hence}, we have calculated the probability of occurrence of specific knot types within the knotted phase for the case of $r_b/r_{\rm{min}}=0.891$. We have examined the knotted structures for both models and found that the maximum number of crossing observed was $8$. Thus, at first we made a list of $D$ values of all the possible knots which have $8$ or less crossings. Using this list (see Table \ref{tab1_5}) we now measure the probability of occurrence of specific knot types from our time series data of the knot parameter $D$. These probabilities for a fixed temperature $T=0.01$ and for six different $\kappa$ within the knotted phase in both models are plotted in Figs.\ \ref{Knot-Prob_N28}(a) and (b), respectively for the two models. For the bead-stick model it shows that for lower values of $\kappa < 4$ different knotted structures $3_1$, $4_1$, $5_1$, and $8_{19}$ are mixed with the unknotted structures. For relatively larger values, viz., $\kappa=4-8$ the unknotted structures vanish 
and $8_{19}$ and $5_1$ emerge as the stable structures, respectively. These plots not only confirm the presence of a rich variety of knots for the bead-stick model with $N=28$ but also indicate that the stable knotted phase lies between $\kappa=4$ and $\kappa=8$. 
\par
The corresponding plots for the bead-spring model in Fig.\ \ref{Knot-Prob_N28}(b) also show a somewhat similar picture. Noticeable again is the fact that the unknotted phase vanishes at a much lower $\kappa=2$ compared to the bead-stick model and continues to remain so until $\kappa=10$ indicating a much wider range of stable knotted phases, a fact also encountered for the $N=14$ polymer. The other noticeable 
feature is that for the bead-spring model the stable knotted structure is $5_1$ which has less crossing than the corresponding $8_{19}$ knot for the bead-stick model. This again could be attributed to the 
presence of $E_{\rm FENE}$ in the nonbonded energy. The presence of the FENE bonds allows the polymers to orient its bonds appropriately and thereby lowering the energy. However, for bead-stick polymers since this is not possible they achieve it by making additional crossings which gives rise to more knotted structures like $8_{19}$. 

\section{Conclusion}\label{conclusion}
We have presented results on the existence of stable knotted phases in semiflexible polymers via extensive replica exchange Monte Carlo simulations of a bead-stick and a 
bead-spring homopolymer model covering the full range of the bending stiffness $\kappa$ via which one can tune the polymer from a completely flexible to a stiff one. 
We speculate that the existence of a knotted phase is dependent on the choice of the ratio $r_b/r_{\rm{min}}$ between the equilibrium bond length $r_b$ and the distance $r_{\rm{min}}$ for the maximum nonbonded contact. Via simple qualitative arguments based on the interplay of the energy gain due to nonbonded contacts and the bending energy penalty, it can be understood that for cases where $r_b/r_{\rm{min}} \ne 1$ the knotted structures are more favorable than the alternative bent structures. 
This was strongly supported by our simulation results for different choices of $r_b/r_{\rm{min}}$ for both models. 
\par
When the results of the two models are compared, the knotted phase in the bead-spring model is much wider than the corresponding range in the bead-stick model. In this regard our results for the bead-spring model can be compared with the results of Seaton \textit{et al.}\ \cite{seaton2013flexible} where they do not mention any existence of knotted structures. This could be due to the use of  $r_b/r_{\rm{min}}=1.0$ coupled with the fact they did not perform their simulation at low enough temperature compared to ours. Similarly, in a study of semiflexible polymer adsorbed on surface no knots were found \cite{austin2017interplay}. This could also be attributed to the fact there also $r_b/r_{\rm{min}}=1.0$ and the lowest simulation temperature ($T=0.1$) was much higher than the one where we found the knotted phase in this work. Thus, it would be worth revisiting this issue on the existence of stable knotted structures in polymers adsorbed on surfaces by tuning $r_b/r_{\rm{min}}$ in the model used. This we take as future endevour.
\par
In conclusion, our results point out that knots are generic phases for semiflexible homopolymers except for a very narrow range of choice of the ratio $r_b/r_{\rm{min}}$ close to unity. This is in contrast with the corresponding results on the existence of knots in proteins which are typically modeled as semiflexible heteropolymer. A deeper insight into heteropolymers reveal that this can be plausible due to the following fact. Hompolymers can have substantial energy gain via nonbonded contacts happening due to several crossings or under passing present 
in a knotted structure. However, for a heteropolymer such energy gain is not guaranteed due to the presence of specific hydrophobic and polar sequences. From this point of view it  would also be worth 
exploring the sequence dependent formation of knotted structures in semiflexible heteropolymer which in turn will throw some light on the existence of knots in proteins. 

\begin{acknowledgments}
We thank Stefan Schnabel for useful discussion. This project was funded by the Deutsche Forschungsgemeinschaft (DFG, German Research Foundation) under Grant Nos.\ JA 483/33-1
and 189\,853\,844--SFB/TRR 102 (project B04), and the 
Deutsch-Franz\"osische Hoch\-schule (DFH-UFA) through the Doctoral College ``$\mathbb{L}^4$'' under Grant No.\ CDFA-02-07. It was further supported by the EU COST programme ``EUTOPIA'' under Grant No.\ CA17139. 
\end{acknowledgments}

\appendix*
\section{Harmonic approximation of the potentials used in Ref.\ \cite{seaton2013flexible}}\label{Appendix1}
The semiflexible polymer model used by Seaton \textit{et al.}\ \cite{seaton2010collapse,seaton2013flexible} is a bead-spring model where the nonbonded interaction is given by a Lennard-Jones (LJ) potential and the bending energy 
penalty was constructed in the same fashion as we did. The main difference is the bond energy. Following we describe the form of potentials they used and subsequently do the harmonic approximation of their bond energy. The nonbonded potential among the monomers is given by
\begin{equation}\label{Seaton_LJ}
E_{\rm{nb}}(r) = \begin{cases}
    E_{\rm{LJ}}(r) - E_{\rm{LJ}}(r_c) ~~~~~~ r<r_c \,,\\
    0 ~~~~~~~~~~~~~~~~~~~~~~~~~~~ \text{otherwise}\,.
  \end{cases}
\end{equation}
where
\begin{equation}
 E_{\rm{LJ}}(r) = \epsilon \left[\left(\frac{\sigma}{r}\right)^{12}- 2\left(\frac{\sigma}{r}\right)^6\right]
\end{equation}
and $r_c=3\sigma$ \cite{seaton2010collapse}. Here, they choose $\sigma=1$ and $\epsilon=1$ such that $r_{\rm{min}}=1$ with $E_{\rm{LJ}}(r_{\rm{min}})=-\epsilon=-1$ and $E_{\rm{LJ}}(r)$ in \eqref{Seaton_LJ} agrees exactly with our form for $E_{\rm{LJ}}(r)$ 
in \eqref{our_LJ} with $\sigma=2^{-1/6}$ albeit the cut-off distance $r_c=3$ is different from our $r_c=2.5\sigma \approx 2.23$.

\par
The bonded interaction between two  monomers consists of a combination of a finitely extensible nonlinear elastic (FENE) and the LJ potential described above,
\begin{equation}\label{Seaton_bond}
E_{\rm{b}}(r) = \begin{cases}
    E_{\rm{FENE}}(r) + E_{\rm{LJ}}(r) ~~~~~~~~ 0 < r \leq R_0,\\
    0 ~~~~~~~~~~~~~~~~~~~~~~~~~~~~~~~~~~~ \text{otherwise}.
  \end{cases}
\end{equation}
The LJ potential in \eqref{Seaton_bond} has the same form as in \eqref{Seaton_LJ} but the values of the parameters are different which will be discussed below. Here, the FENE potential has the form \cite{kremer1990dynamics}
\begin{equation}
 E_{\rm{FENE}}(r) = - \frac{K}{2}R_0^2 {\rm{ln}} \left[ 1- \left(\frac{r}{R_0}\right)^2\right],
\end{equation}
where $R_0$ is the finite extensibility and $K$ is the stiffness constant. In dimensionless units, 
the values were taken as $R_0=1.2$ and $K=2$ by Seaton \textit{et al.}. Note that their choice of $E_{\rm{FENE}}$ is different from $E_{\rm{FENE}}$ we have chosen for our simulations as given in Eq.\ \eqref{FENE}.

\par
They determined the parameters of $E_{\rm{LJ}}$ in such a way that $E_b$ is minimum at bond length $r=r_{b}=1$. Thus, setting the first derivative of this bonded potential $\left(\frac{d E_b}{dr}\right)=0$ gives
\begin{equation}
 \frac{12 \epsilon}{r_{b}} \left[\left(\frac{\sigma}{r_{{{b}}}}\right)^{12}- \left(\frac{\sigma}{{r_{{b}}}}\right)^6\right] = \frac{K r_{{{b}}}}{1-(r_{{{b}}}/R_0)^2}.
\end{equation}
Now solving this equation with $r_{b}=1$ gives us a dependence of $\sigma$ on $\epsilon$ as (using $R_0=1.2$ and $K=2$), 
\begin{equation}
 \sigma^6 = \frac{1}{2}\left(1+\sqrt{1+\frac{24}{11\epsilon}}\right).
\end{equation}
By setting $\epsilon=2$ \footnote{We thank D. Seaton and S. Schnabel for providing the parameters they used in Ref.\ \cite{seaton2013flexible}.} in this equation, one gets $\sigma \approx 1.03412$. 

\par
In order to obtain the effective spring constant for the bonded potential, we need to do a Taylor series 
expansion of $E_{\rm{b}}(r)$ around its minimum ($r_{b}$) and keep the terms up to the second order derivative. The expansion gives 
\begin{equation}
 E_{\rm{b}}(r) = E_{\rm{b}}(r_b)+\left. (r-r_b) \frac{dE_{\rm{b}}}{dr}\right|_{r=r_b } + \left.\frac{(r-r_b)^2}{2} \frac{d^2E_{\rm{b}}}{dr^2}\right|_{r=r_b}+ \dots~.
\end{equation}
Keeping up to the harmonic approximation (i.e., up to the second derivative) and shifting $E_{\rm{b}}(r)$ by $E_{{\rm b}}(r_b)$ we get,
\begin{equation}
 E_{\rm{b}}(r) = \frac{K_{\rm{eff}}}{2} (r-r_b)^2,
\end{equation}
where 
\begin{equation}
 K_{\rm{eff}} = \left.\frac{d^2E_{\rm{LJ}}}{dr^2}\right|_{r=r_{{b}}} + \left.\frac{d^2E_{\rm{FENE}}}{dr^2}\right|_{r=r_{b}}.
\end{equation}

\par
\begin{figure}[t!]
\includegraphics[width=0.4 \textwidth]{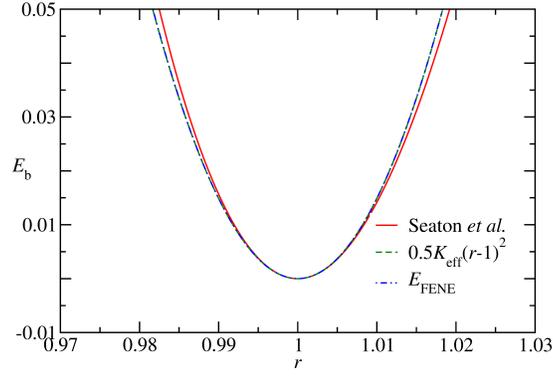}
\caption{Comparison of the different bond potential employed by Seaton \textit{et al.}\ \cite{seaton2013flexible,seaton2010collapse}, a simple harmonic potential with spring constant $K_{\rm {eff}}=297.5$, and the FENE potential used by us as 
in Eq.\ \eqref{FENE} with $K=K_{\rm {eff}}$.}
\label{harmonic}
\end{figure}

Now using the second-order derivatives of both the terms for the bonded potential, we get $K_{\rm{eff}}$ (for $r_{b}=1$) as,
\begin{equation}\label{K_eff}
 K_{\rm{eff}} = 12 \epsilon \left[13 \sigma^{12}- 7\sigma^{6}\right] + \frac{KR_0^2}{(R_0^2-1)^2} (R_0^2+1).
\end{equation}
Inserting the values of the parameters ($\epsilon=2$, $\sigma=1.03412$, $R_0=1.2$ and $K=2$) in the above equation gives the effective value of the 
spring constant as $ K_{\rm{eff}} \approx 297.5$.
{As can be seen in Fig.\ \ref{harmonic}, for small variations of the bond length the agreement is excellent.}
\begin{figure}[b!]
\includegraphics[width=0.4 \textwidth]{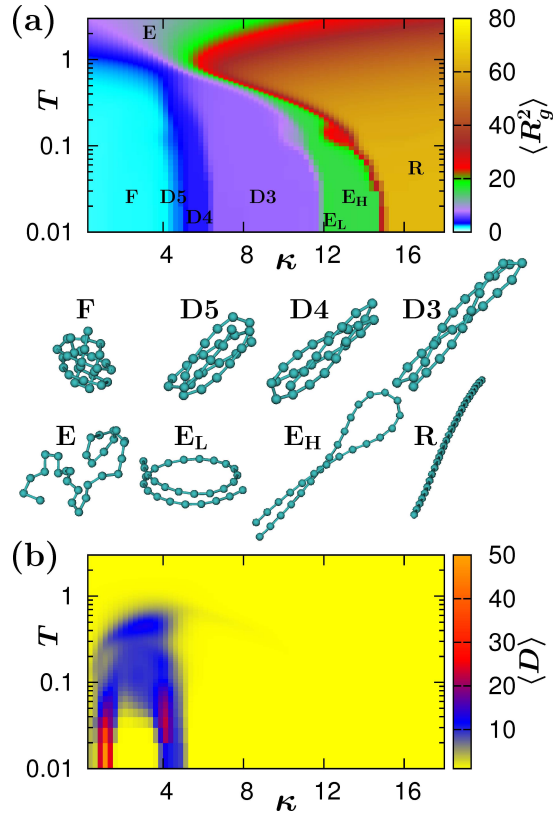}
	\caption{(a) Complete phase diagram for a semiflexible polymer of length $N=28$ using the bead-spring model with $r_b/r_{\rm{min}}=1.0$ and $K=K_{\rm{eff}}=297.5$. The surface plot is generated using the estimated squared radius of gyration $\langle R_g^2 \rangle$. The labeled phases stand for the following: E for elongated; R for rodlike; F for frozen;  D$n$ for bent phases with $n$ number of segments; $\rm{E_L}$ for elongated loop; $\rm{E_H}$ for hairpin. (b) The corresponding phase diagram using the knot parameter $\langle D \rangle$ as the order parameter.}
 \label{Seaton_PD}
\end{figure}
\par
Using $K=K_{\rm{eff}}$ in our bead-spring model with $r_b/r_{\rm{min}}=1.0$ we now perform replica exchange simulations. The results are summarized in Fig.\ \ref{Seaton_PD} for a polymer with a choice of $N=28$, consistent with the largest choice we made in the main text. Nevertheless, $N=28$ is almost as long as in Ref.\ \cite{seaton2013flexible} where $N=30$ was chosen. The phase diagram in Fig.\ \ref{Seaton_PD}(a) shows that we also observe the same variety of conformations as was obtained in Ref.\ \cite{seaton2013flexible}, that includes frozen (F), extended coil (E), bent structures (D5, D4, D3), elongated loop ($\rm{E_L}$), hairpin ($\rm{E_H}$), and rod-like (R) structures. On the other hand, the phase diagram with the estimated knot parameter $\langle D \rangle$ as the order parameter in Fig.\ \ref{Seaton_PD}(b) shows no stable knotted phase. The blue region in there is only an indication of mixed phases that constitute  unknotted frozen structures and  knotted structures which we confirmed from the corresponding time series of $D$.
 \section*{References}
%
\end{document}